\documentstyle[12pt]{article}

\begin{document}
\newcommand{\ice}[1]{\relax}
\newcommand{\be}{\begin{equation}}
\newcommand{\ee}{\end{equation}}
\newcommand{\al}{\alpha}
\newcommand{\G}{\Gamma}

\title{
\vspace*{-40mm}
\begin{flushright}
{\large \bf  INR-98-0986\\[-5mm]
July 1998}\\[5mm]
\end{flushright}   
{
\bf Bottom quark  mass and 
$|V_{cb}|$ matrix element 
from  $R(e^+e^-\rightarrow b\bar b)$ and 
$\Gamma_{\rm sl}(b\rightarrow cl\nu_l)$ 
in the next-to-next-to-leading order.
}}
\author{
  {\bf A.A.Penin and A.A.Pivovarov}\\
  {\small {\em Institute for Nuclear Research of the
  Russian Academy of Sciences,}}\\
  {\small {\em 60th October Anniversary
  Pr., 7a, Moscow 117312, Russia}}
        }

\date{}

\maketitle

\begin{abstract}
We present a consistent analysis of $\Upsilon$ sum rules
and  $B$-meson semileptonic width in the 
next-to-next-to-leading order in  the strong coupling constant.
The analysis is based on the analytical result for  
the heavy quark vector current correlator
near threshold in the second order in perturbative and relativistic 
expansion around the nonrelativistic Coulomb approximation.
\\[2mm]
\end{abstract}

\thispagestyle{empty}
\newpage

\section{Introduction.}
The rich phenomenology of the heavy quark mesons has been 
recognized as a  source of information on strong 
interaction dynamics and a clean place to determine 
fundamental parameters of the standard model such as the strong coupling
constant, heavy quark masses and mixing angles. 
The necessity of the quantitative analysis of 
the strong interaction effects in the heavy flavor physics
leads to developing  new theoretical methods such as  
nonrelativistic QCD (NRQCD) \cite{CasLep} for description of heavy quarkonium
and heavy quark effective theory (HQET) \cite{HQET} for description of 
heavy-light mesons. One of the most promising applications of these
techniques to the bottom quark physics is 
the analysis of the  $\Upsilon$ sum rules \cite{NSVZ}  
and $B$-meson semileptonic decays (see \cite{N,U} for a recent review).
The strict estimate of the actual 
precision of theoretical predictions is now becoming 
important since the experimental precision is rather high and 
is quantitatively comparable with the last available
terms of perturbative approximation. 
Recently a significant progress has been achieved
in calculations of the high order corrections for
the  $\Upsilon$ sum rules \cite{Vol,KPP,PP,H,MY} 
and inclusive semileptonic width of  
$B$-mesons \cite{BLM,CM,PP2}.  

In this paper we present a simultaneous analysis of
$\Upsilon$ sum rules and the $B$-meson inclusive semileptonic 
width $\Gamma_{\rm sl}$ in the next-to-next-to-leading order (NNLO).
On the basis of this analysis we determine the mass 
parameter of the bottom quark in the second order of perturbation
theory\footnote{
Because quark states have not been observed as asymptotic free 
particle states the notion of their mass can not be unambiguously defined.
There are different ways to reflect the fact the $b$-quark is heavy
by introducing its mass \cite{Tarrach}.  
In the present paper we use the notion of the $b$-quark pole mass
defined strictly within the finite order perturbation theory.
Within this definition this mass parameter is equivalent (and can be
reexpressed
through a finite series in $\al_s$) to any other perturbative
definition 
of the quark mass parameter, in particularly, $\overline{\rm MS}$
mass 
in every order of the perturbation theory as 
a parameter of perturbative QCD.}
and give a new accurate estimate of 
Cabibbo-Kobayashi-Maskawa (CKM) matrix element $|V_{cb}|$.

The results of the analysis of the $\Upsilon$ system in 
the NNLO 
approximation in the coupling constant and
nonrelativistic expansion have been briefly reported in ref.~\cite{PP}.
Then two more papers with NNLO analysis of the $\Upsilon$ system 
appeared \cite{H,MY}.
Since there is some difference between the analyses of
ref.~\cite{PP} and refs.~\cite{H,MY} in some details of the 
treatment of
the problem, 
in this paper we give more extensive discussion paying 
the main attention to the self-consistence of the approach, 
uncertainties of theoretical estimates and 
the reliability of results. We discuss also the 
general structure and the asymptotic
character of perturbative expansion for the moments 
of the $\Upsilon$ system spectral density. 
By studying the behavior of the expansion
around the Coulomb approximation for different moments
we estimate the critical order where the asymptotic growth
of the coefficients of the series in $\al_s$ 
for the moments starts.  

The obtained  moments of the $\Upsilon$ system spectral
density are related then to the inclusive 
$B$-meson semileptonic width up to the  
NNLO order. In this way we avoid 
the strong dependence of $\Gamma_{\rm sl}$
on $m_b$ and reduce
the theoretical uncertainty in  $|V_{cb}|$.
In particular, the obtained perturbative expansion 
for $V_{cb}$ converges well (in heuristic sense) up to NNLO in $\al_s$. 

The paper is organized as follows.
In the next Section we describe the vacuum polarization function of heavy
quarks near threshold in the NNLO.
In Section \ref{secsumr} we discuss the $\Upsilon$ sum rules and 
numerical estimates for the $b$-quark
pole and $\overline{\rm MS}$ mass within perturbation theory. 
In Section \ref{inclusive} 
the inclusive $B$-meson semileptonic width and  $|V_{cb}|$
matrix element are determined. The last Section 
contains our conclusions.
In Appendix we give an explicit analytical formula for the polarization
function of heavy quarks near threshold. 

\section{Next-to-next-to-leading order heavy quark  
vacuum  polarization function near threshold.} 
We study the near 
threshold behavior of the  vacuum polarization function  $\Pi(s)$
of the $b$-quark vector current $j_\mu=\bar b\gamma_\mu b$
\[
\left(q_\mu q_\nu-g_{\mu \nu}q^2\right)\Pi(q^2)=
i\int dxe^{iqx}\langle 0|Tj_{\mu}(x)j_{\nu}(0)|0\rangle \,.
\]
The finite order perturbation theory expansion  
in the strong coupling constant $\alpha_s$ is known to  
break down near the two-particle production threshold, $s\sim 4m_b^2$
where $m_b$ is the $b$-quark pole mass.
However, near the threshold, {\it i.e.} for a small $b$-quark velocity 
$v=\sqrt{1-4m_b^2/s}\ll 1$, 
the nonrelativistic approximation becomes valid \cite{CasLep}. 
In the NNLO the nonrelativistic expansion for 
the polarization function  $\Pi(s)$
has the form
\be
\Pi(s)
={N_c\over 2 m_b^2}\left(
C_h(\al_s)G(0,0,k)+{4\over 3}{k^2\over m_b^2}G_C(0,0,k)\right)
\label{thmom}
\ee
with $k=\sqrt{m_b^2-s/4}$ being a natural energy variable near
threshold. The quantity $C_h(\al_s)$ is a
perturbative coefficient that matches
correlators of relativistic and nonrelativistic vector currents
and accounts for the hard part of the QCD corrections. 
It is now known to the second order in $\al_s$ expansion
\[
C_h(\al_s)
=1-C_h^1C_F{\al_s\over \pi}+C_h^2C_F\left({\al_s\over \pi}\right)^2
\]
with 
$
C_h^1=4$ \cite{KS}
and 
\[
C_h^2=\left({39\over 4}-\zeta(3)+{4\pi^2\over 3} 
\ln{2}-{35\pi^2\over 18}\right)C_F-\left({151\over 36}+{13\over 2}
\zeta(3)+{8\pi^2\over 3} \ln{2}-{179\pi^2\over 72}\right)C_A
\]
\be
+
\left({44\over 9}-{4\pi^2\over 9}+{11\over 9}n_f\right)T_F
+2\beta_0\ln{\left(m_b\over\mu\right)}+\pi^2\left({2\over 3}C_F+C_A
\right)\ln{\left(m_b\over\mu_f\right)}
\label{ch}
\ee
where $\al_s$ is defined in $\overline{\rm MS}$ 
renormalization scheme 
with the scale parameter $\mu$. Another scale parameter
$\mu_f$ is a factorization scale which separates
contributions coming from the hard and soft momentum regions
and plays the role of an infrared cutoff in the
diagrams contributing to the quantity $C_h(\al_s)$ \cite{Hoang,Mel}. 
The color symmetry 
$SU(3)$ group invariants for QCD are $C_A=3$, $C_F=4/3$, $T_F=1/2$,
$n_f$ is the number of light fermion flavors, and
$
\beta_0=11C_A/3-4T_Fn_f/3$ is the first $\beta$-function
coefficient.
Here $\gamma_E=0.577216\ldots$ is the Euler constant and $\zeta(z)$ is 
the Riemann $\zeta$-function. 
The quantity $G({\bf x},{\bf y},k)$ is the nonrelativistic 
Green function. This Green function sums up the singular 
threshold corrections 
and satisfies the following Schr{\"o}dinger equation
\[
\left(-{{\bf \Delta}_{\bf x}\over m_b}-{{\bf \Delta}_{\bf x}^2\over 4m_b^3}
+V_C(x)+{\al_s\over 4\pi}\Delta_1V(x)+\left({\al_s\over 4\pi}\right)^2
\Delta_2V(x)\right.
\]
\be
\left.
+\Delta_{NA}V(x)+\Delta_{BF}V({\bf x},\partial_{\bf x},{\bf S}) 
+ {k^2\over m_b}\right)
G({\bf x},{\bf y},k)=\delta({\bf x}-{\bf y})
\label{Schr}
\ee
where ${\bf \Delta_x}= \partial_{\bf x}^2$, $V_C(x)=-C_F\al_s/x$ is the  Coulomb
potential, $x=|{\bf x}|$,
$\Delta_{NA}V(x)=-C_AC_F\al_s^2/(2m_bx^2)$ is the so called non-Abelian potential 
of quark-antiquark interaction \cite{Gup}, 
\linebreak
$\Delta_{BF}V({\bf x},
\partial_{\bf x},{\bf S})$ 
is the standard Breit-Fermi potential (up to the color factor $C_F$)
containing the quark 
spin operator ${\bf S}$, {\it e.g.}~\cite{Landau}.
The terms $\Delta_iV$ ($i=1,2$) represent
the first and second order perturbative QCD corrections to the  Coulomb
potential \cite{Fish,Peter,YS}\footnote{The value of the
$a_2$ coefficient in eq.~(\ref{potcorr1}) obtained in
ref.~\cite{Peter} exceeds the correct
result obtained recently \cite{YS} by $2\pi^2 C_A^2$. Though the
value of ref.~\cite{Peter} was used in the previous
analysis of the $\Upsilon$ sum rules \cite{PP} and $B$-meson
semileptonic width \cite{PP2} its difference
from the present one is numerically small and results in no significant
change for the numerical estimates of $m_b$, $\al_s$
and $|V_{cb}|$  given in these papers.
In the present paper the most recent value of ref. \cite{YS}
is used.}        
$$
\Delta_1V(x)={\al_s\over 4\pi}V_C(x)(C_0^1+C_1^1\ln(x\mu)),
$$
\be
\Delta_2V(x)=\left({\al_s\over 4\pi}\right)^2V_C(x)(C_0^2+C_1^2\ln(x\mu)+C_2^2 \ln^2(x\mu))
\label{potcorr1}
\ee
where  
\[
C_0^1=a_1+2\beta_0\gamma_E,\qquad C_1^1=2\beta_0,
\]
\[
C_0^2=\left({\pi^2\over 3}+4\gamma_E^2\right)\beta_0^2
+2(\beta_1+2\beta_0a_1)\gamma_E+a_2,
\]
\[
C_1^2=2(\beta_1+2\beta_0a_1)+8\beta_0^2\gamma_E,
\qquad 
C_2^2=4\beta_0^2,
\]
$$
a_1={31\over 9}C_A-{20\over 9}T_Fn_f,
$$
\[
a_2= \left({4343\over 162}+4\pi^2-{\pi^4\over 4}
+{22\over3}\zeta(3)\right)C_A^2-
\left({1798\over 81} + {56\over 3}\zeta(3)\right)C_AT_Fn_f
\]
$$
-\left({55\over 3} - 16\zeta(3)\right)C_FT_Fn_f
+\left({20\over 9}T_Fn_f\right)^2,
$$
\[
\beta_1={34\over 3}C_A^2-{20\over 3}C_AT_Fn_f-4C_FT_Fn_f.
\]
The second term in eq.~(\ref{thmom}) is generated by the 
operator of dimension five in  
the  nonrelativistic expansion of the 
vector current (see, {\it e.g.} \cite{Bod}). It contains 
the Green function of the pure Coulomb Schr{\"o}dinger equation
at the origin $G_C({\bf x},{\bf y},k)|_{x,y=0}$ \cite{Schw}. 
In the short distance limit $x\rightarrow 0$
the Coulomb Green function $G_C({\bf x},0,k)$ has $1/x$ and $\ln x$
divergent terms. These terms, however, have no imaginary 
part and do not contribute
to the spectral density of the polarization function.
Hence they can be subtracted. After the subtraction
the (renormalized) Coulomb Green function takes the form   
\be
G^r_C(0,0,k)=-{C_F\al_sm_b^2\over 4\pi}\left({k\over C_F\al_sm_b}+
\ln\left({k\over \mu_f}\right)+\gamma_E+\Psi_1\left(1- {C_F\al_sm_b\over 2k}
\right)\right)
\label{G0}
\ee
where $\Psi_n(z)=d^n\ln{\Gamma(z)}/dz^n$ and
$\Gamma(z)$ is the Euler $\Gamma$-function.
   
The solution of eq.~(\ref{Schr}) can be 
found within the standard nonrelativistic 
perturbation theory
around the Coulomb Green function as a leading order approximation
\[
G(0,0,k)=G_C(0,0,k)+\Delta G(0,0,k),
\]
\be
\Delta G(0,0,k)=-\int G_C(0,{\bf x},k)\left(
-{{\bf \Delta}_{\bf x}^2\over 4m_b^3}+
{\al_s\over 4\pi}\Delta_1V({x})+
\ldots\right)G_C({\bf x},0,k)d{\bf x}+\ldots
\label{totcorr}
\ee
\[
=\Delta_{{\bf \Delta}^2,NA,BF}G+\Delta_1G+\Delta_2^{(2)}G+
\Delta_2^{(1)}G+\ldots 
\]
The corrections to the Coulomb Green function at the origin
due to 
${\bf\Delta}^2$, $V_{NA}$ and  $V_{BF}$ terms are known 
analytically \cite{Hoang,Mel}.
The nontrivial part of the calculation consists in a proper treatment 
of some integrals in eq.~(\ref{totcorr}) that correspond to 
these corrections and diverge, or become ill-defined at small $x$. 
This divergence is a consequence of the fact that the nonrelativistic 
approximation is not relevant for the description of the
short distance effects. 
Moreover, in contrast to the next-to-leading order 
there exists a divergence in the imaginary part
of the Green function in NNLO that contributes to the spectral density.
The divergence can be regularized by introducing an 
ultraviolet cutoff $\mu_f$. Then, following the general 
line of the effective field theory approach one has  
to match the calculation of the Green function to the 
calculation of the coefficient $C_h(\al_s)$ 
{\it i.e.} to match the regularization procedures for 
the $G(0,0,k)$ and  $C_h$.
This can be done by comparing eq.~(\ref{thmom}) and 
the result of perturbative calculation of the spectral density \cite{sd}
in the formal limit $\al_s\ll v\ll 1$   up to the order $\al_s^2$ 
\cite{Hoang,Mel}. The  result reads\footnote{Calculating  
the NNLO  corrections to the nonrelativistic Green function and 
the hard renormalization coefficient one encounters 
the divergences  with the specific form depending
on the regularization procedure. They are not included
to eqs.~(\ref{ch},~\ref{Hoangcorr}) because the divergent terms
in the hard renormalization coefficient cancel the divergence 
in the corrections to the nonrelativistic Green function  leaving the 
logarithmic dependence of eq.~(\ref{ch}) on the cutoff $\mu_f$
which compensates corresponding dependence of eq.~(\ref{Hoangcorr}).}
\[
\Delta_{{\bf \Delta}^2,NA,BF}G=
{C_F\al_sm_b^2\over 4\pi}\left(
{5\over 8}{k^3\over C_F\al_sm_b^3}+
2{k^2\over m_b^2}\left(
\ln\left({k\over \mu_f}\right)\right.\right.
\]
\be
\left.\left.
+\gamma_E
+\Psi_1\left(1- 
{C_F\al_sm_b\over 2k}\right)\right)
-{11\over 16}{C_F\al_s k\over m_b}
\Psi_2\left(1-{C_F\al_sm_b\over 2k}\right)
\right)
\label{Hoangcorr}
\ee
\[
+{4\pi\over 3}{C_F\al_s\over m_b^2}
\left(1+{3\over 2}{C_A\over C_F}\right)G^r_C(0,0,k)^2 .
\]
The next-to-leading (NLO) correction $\Delta_1G$ in eq.~(\ref{totcorr}) 
due to the first iteration
of $\Delta_1V$ term of the QCD potential 
and  the NNLO correction $\Delta_2^{(2)}G$ 
due to  $\Delta_2V$ part of the potential 
have been found in ref.~\cite{KPP}. 
The correction $\Delta_2^{(1)}G$ due to the 
second iteration of $\Delta_1V$ term which has to 
be kept in NNLO approximation 
has been obtained in ref.~\cite{PP}.
We  describe the details of this calculation below. 
 
We use the following partial wave representation for the 
Coulomb Green function
\[
G_C({\bf x},{\bf y},k)=\sum^\infty_{l=0}(2l+1)G_l(x,y,k)P_l(({\bf xy})/xy)
\]
\be
G_l(x,y,k)={m_bk\over 2\pi}(2kx)^l(2ky)^le^{-k(x+y)}
\sum_{m=0}^\infty {L_m^{2l+1}(2kx) L_m^{2l+1}(2ky)m!\over
(m+l+1-\alpha_sC_Fm_b/(2k))(m+2l+1)!}
\label{rep}
\ee
where $P_l(z)$ is a Legendre polynomial and $L^\al_m(z)$ is a Laguerre  polynomial
\[
L_m^\al(z)={e^zz^{-\al}\over m!}\left({d\over dz}\right)^m
(e^{-z}z^{m+\al}).
\]
Only $l=0$ component of eq.~(\ref{rep}) is necessary 
for the calculation of the corrections to the   Green function
at the origin up to  NNLO approximation.

Let us consider first the NLO correction.
It can be written as
\be
\Delta_1G={\al_s\over 4\pi}\left({m_bk\over 2\pi}\right)^2
\sum_{m,n=0}^\infty H(m)H(n)\int e^{-2kx}L_m^1(2kx)L_n^1(2kx)
\Delta_1V(x)d{\bf x}
\label{G1a}
\ee
where 
$$
H(m)=\left(m+1-{\displaystyle {C_F\al_s m_b\over 2k}}\right)^{-1} 
$$ 
and we use the equality
\[
L_m^\al(0)={\Gamma(m+\al+1)\over \Gamma(\al+1)\Gamma(m+1)}.
\]
The  integrals in eq.~(\ref{totcorr}) corresponding to 
$\Delta_1V$ term also diverge at small $x$.
The divergent part however is $k$ 
independent and does not 
contribute to the spectral density. 
As a consequence no matching is necessary for 
calculation of these corrections.  
In the representation~(\ref{G1a}) the divergence of the integral
at small $x$ is transformed to the divergence of the sum.  
The divergent part of eq.~(\ref{G1a})   
can be separated by the following method.
Eq.~(\ref{G1a}) can be rewritten in the form
\[
\Delta_1G={\al_s\over 4\pi}\left({m_bk\over 2\pi}\right)^2\left(
\sum_{m,n=0}^\infty F(m)F(n)\int e^{-2kx} 
L_m^1(2kx)L_n^1(2kx)\Delta_1V(x)d{\bf x}
\right.
\]
\be
\left.
+2\sum_{m=0}^\infty F(m)\int 
{e^{-2kx}L_m^1(2kx)\Delta_1V(x)\over 2kx}d{\bf x}
+\int {e^{-2kx}\Delta_1V(x)\over (2kx)^2}d{\bf x}\right)
\label{G1b}
\ee
where 
\[
F(m)={C_F\al_s m_b\over (m+1)2k}\left(m+1-{ 
{C_F\al_s m_b\over 2k}}\right)^{-1} 
\] 
and we used the property of the  Laguerre  polynomial
\[
\sum_{m=0}^\infty{L_m^\al(z)\over m+\al}=z^{-\al}\Gamma(\al).
\]
In  eq.~(\ref{G1b}) all sums are convergent and the divergence
is contained in the last term. 
Two divergent integrals in the last term of this equation
after a regularization take the form 
\[
\int {e^{-2kx}\ln(\mu x)\over x}dx=
-\gamma_EL(k)+{1\over 2}L(k)^2+\ldots ,
\]
\[
\int {e^{-2kx}\over x}dx=L(k)+\ldots
\]
where $L(k)=-\ln(2k/\mu)$ and 
ellipsis stands for inessential $k$ independent divergent parts.
Two finite integrals in eq.~(\ref{G1a}) are explicitly given by the
following expression
\[
\int e^{-z}L^1_n(z)L^1_m(z)\ln(z)zdz=
\left\{
\begin{array}{c}
(m+1)\Psi_1(m+2), \quad m=n, \\
{\displaystyle -{n+1\over m-n}}, \quad m>n, \\
\end{array} 
\right. 
\]
\be
\int e^{-z}L^1_m(z)\ln(z)dz= -2\gamma_E-\Psi_1(m+1).
\label{integ1}
\ee
To compute the first integral we rewrite it in the form
containing a derivative with respect to an auxiliary parameter 
$\varepsilon$ 
\be 
\left.
\int e^{-z}L^1_n(z)L^1_m(z)\ln(z)zdz=
{d\over d\varepsilon}\left(\int e^{-z}L^1_n(z)L^1_m(z) 
z^{1+\varepsilon}dz
\right)\right|_{\varepsilon =0}.
\label{integ2}
\ee
By using the relations 
\[
L^\beta_m(z)=\sum_{n=0}^m{\Gamma(\beta-\al +n)\over\Gamma(\beta-
\al)\Gamma(n+1)}L_{m-n}^\al(z),
\]
\[ 
\int e^{-z}L^\al_n(z)L^\al_m(z)z^\al dz=
\delta_{mn}{\Gamma(m+\al +1)\over\Gamma(m+1)}
\]
for $\beta=1$, $\al=1+\varepsilon$ the integration in
the right hand side of eq.~(\ref{integ2}) can be performed
analytically.
Then taking the derivative in $\varepsilon$ at 
$\varepsilon=0$ we get the first line of eq.~(\ref{integ1}).  
The second integral in eq.~(\ref{integ1}) can be computed 
using the same technique.  
Thus the final result for the correction is
$$
\Delta_1G={\al_s\over 4\pi}{C_F\al_sm_b^2\over 4\pi}\left(
\sum_{m=0}^\infty F^2(m)(m+1)
\left(C_0^1+(L(k)+\Psi_1(m+2))C_1^1\right)
\right.
$$
\[
-2\sum_{m=1}^\infty\sum_{n=0}^{m-1}
F(m)F(n)
{n+1\over m-n}C_1^1 
+2\sum_{m=0}^\infty F(m)
\left(C_0^1+(L(k) -
2\gamma_E-\Psi_1(m+1))C_1^1\right)
\]
\be
\left.+L(k)C_0^1+\left(-\gamma_E L(k)+
{1\over 2}L(k)^2\right)C_1^1
\right).
\label{G1}
\ee
Note that the nontrivial part of the calculation 
is to find the correction to the Green function due to
the logarithmic correction to the Coulomb potential~(\ref{potcorr1}). 
Computation of the 
correction due to the constant part of the correction to 
the Coulomb potential is trivial. 
It can be also found from the leading
order result by changing the parameter of pure
Coulomb solution $\al_s \rightarrow \al_s(1+C_0^1\al_s/4\pi)$.

The $\Delta_2^{(2)}G$ correction has been obtained in the same way.
The  $\Delta_2^{(1)}G$ part is finite and  requires no
regularization. It can be computed directly using  
the representation~(\ref{rep}) and the first integral
of eq.~(\ref{integ1}).
The results of the calculations are given in Appendix. 

The  Green function at the origin 
can be written in the  form which includes 
only single poles in the energy variable. Such a form looks more natural
for the Green function of a nonrelativistic Schr{\"o}dinger equation
\be
G(0,0,E)=\sum_{m=0}^\infty{|\psi_m(0)|^2\over E_m-E}+
{1\over \pi}\int_0^\infty{|\psi_{E'}(0)|^2\over E'-E}
dE'
\label{endenom}
\ee
where $E=-k^2/m_b$, $\psi_{m,E'}(0)$ is the wave function at the origin, 
the sum goes over bound states 
and the integral is over the state of continuous  part of the
spectrum.
In this way the corrections to the Green function stemming from
the discrete part of the spectrum 
reduce to
corrections to the Coulomb bound state energy levels
\be
E_m=-{C_F^2\al_s^2m_b\over 4(m+1)^2}\left(1+\Delta_1E_m
+\Delta_{{\bf \Delta}^2,NA,BF}E_m+\Delta^{(2)}_2E_m
+\Delta^{(1)}_2E_m\right)
\label{ecorr}
\ee
and to the values of Coulomb bound state wave functions at the origin 
\be
|\psi_m(0)|^2={C_F^3\al_s^3m_b^3\over 8\pi(m+1)^3}\left(1+\Delta_1\psi^2_m
+\Delta_{d5}\psi^2_m+\Delta_{{\bf\Delta}^2,NA,BF}\psi^2_m+\Delta^{(2)}_2\psi^2_m
+\Delta^{(1)}_2\psi^2_m\right)
\label{pscorr}
\ee
where $\Delta_{d5}\psi^2_m$ is the correction due to the second term in 
eq.~(\ref{thmom}).

In NLO an explicit analytical expression for the corrections
to the bound state parameters has the form
\be
\Delta_1E_m={\alpha_s\over 4\pi}\left(2C_0^1+2(L(m)+
\Psi_1(m+2))C_1^1\right),
\label{e1}
\ee
\[
\Delta_1\psi^2_m={\alpha_s\over 4\pi}\left( 3C^1_0+
(3L(m)-1-2\gamma_E+{2\over m+1}+\Psi_1(m+2)
-2(m+1)\Psi_2(m+1))C_1^1\right)
\]
where 
\[
L(m)=\ln\left({(m+1)\mu\over C_F\al_sm_b}\right).
\]
The expressions of the NNLO
corrections are rather cumbersome and given in
Appendix.

The principal possibility of 
relating results of eqs.~(\ref{ecorr},~\ref{pscorr})
to the mass and leptonic width of
corresponding $\Upsilon$ resonances 
with an account for nonperturbative corrections was discussed in \cite{VL}.

Thus we have described some details of obtaining 
the complete analytical expressions for the vacuum polarization
function of heavy quarks near the two-particle 
threshold in NNLO presented in \cite{KPP,PP}
\footnote{The NNLO vacuum polarization
function of heavy quarks was also recently obtained
in a different representation in ref.~\cite{MY} 
where the ``low scale'' mass of $b$-quark was studied
in the context of $\Upsilon$ sum rules.}. 
Note that our result does not determine the 
additive renormalization constant in the real part
of the  polarization function which has to be
fixed according to the standard normalization condition 
$\Pi (0)=0$. This constant, however, is 
inessential for physical applications.

\section{$\Upsilon$ sum rules and $b$-quark
mass}
\label{secsumr} 
The result for the near
threshold behavior of the vacuum polarization function of the heavy
quark is now applied to the analysis 
of sum rules for the $\Upsilon$ system.
Within the sum rules approach the  moments ${\cal M}_n^{th}$ 
\[
{\cal M}_n^{th}= 
\left.{12\pi^2\over n!}(4m_b^2)^n{d^n\over ds^n}\Pi(s)\right|_{s=0}=
(4m_b^2)^n\int_0^\infty{R(s)ds\over s^{n+1}}
\]
of the spectral density 
$R(s)=12\pi {\rm Im}\Pi(s+i\epsilon)$
of the theoretical vacuum polarization function of the heavy
quarks $\Pi(s)$  should be compared with the experimental ones 
\be
\label{expmom0}
{\cal M}_n^{exp} = 
{(4m_b^2)^n\over Q_b^2}\int_0^\infty{R_b(s)ds\over s^{n+1}}
\ee
under the assumption of quark-hadron duality.
Here $Q_b=-1/3$ is the $b$-quark electric charge
and the theoretical parameter $m_b$ is included for convenience.
The experimental moments ${\cal M}_n^{exp}$ are generated by 
the function $R_b(s)$ which is the 
normalized cross section of $e^+e^-$ annihilation
$R_b(s) = \sigma(e^+e^-\rightarrow {\rm hadrons}_{\,b\bar b})/
\sigma(e^+e^-\rightarrow \mu^+\mu^-)$ and, at least in principle, can
be directly found from experiment for any $s$. 
In practice, however, the spectral density is well measured from 
experiments only for
values of energy rather close to threshold where the well
pronounced $\Upsilon$ resonances exist.  
Therefore numerical values for the experimental moments
are obtained  by saturating 
the integral in eq.~(\ref{expmom0}) with the contribution
of
the first six $\Upsilon$ resonances 
while the tail of the spectral density at large energy
is approximated theoretically
\be
{\cal M}_n^{exp}={(4m_b^2)^n\over Q_b^2}
\left({9\pi\over \al_{QED}^2(m_b)}
\sum_{k=1}^6{\Gamma_{k}\over M_{k}^{2n+1}}
+\int_{s_0}^\infty\!{\rm d}s {R_b(s)\over s^{n+1}}\right).
\label{expmom}
\ee
The leptonic widths
$\Gamma_{k}$ and masses $M_{k}$ $(k=1\ldots 6)$ of the resonances
are known with good accuracy \cite{PDG}.
For example, for large $n$ the dominant contribution to the moments
comes from the first
$\Upsilon$ resonance for which 
$\Gamma_{\Upsilon(1S)}=1.32\pm 0.05~{\rm keV}$ and 
$M_{\Upsilon(1S)}=19460.37\pm 0.21~{\rm MeV}$. 
The electromagnetic coupling constant is renormalized to the 
energy of order $m_b$ with the result 
$\al_{QED}^2(m_b)=1.07 \al^2$.
The rest of the spectrum beyond the resonance
region  for energies larger than
$s_0\approx (11.2~{\rm GeV})^2$ (continuum contribution)
is approximated by the theoretical spectral density
multiplied by the parameter $0.5<t<1.5$ which accounts
for the uncertainty in the experimental data in this energy 
region. 
The allowed range for the parameter $t$ is chosen to be 
sufficiently large that gives 
a very conservative estimate of uncertainty caused by all other 
contributions but the resonance one. For instance,
the numerical change of the experimental moments
produced by a manifest account for the physical continuous spectrum due 
to the open $B\bar B$ production
at $s>(10.6~{\rm GeV})^2$ is well within the error bars
introduced by variation of the $t$ parameter.

The ordinary perturbative expansion
in the coupling constant taken up to a finite order 
is not accurate enough 
for sufficiently high moments since the coefficients of 
the expansion in low orders 
of perturbation theory grow fast with the
order of the moment. Therefore a kind of resummation of the terms 
giving a dominant contribution is required.
These terms are identified as those connected with a strong Coulomb
interaction in the final state of quark-antiquark production.
The importance of the Coulomb resummation for large $n$
can by directly inferred from the explicit 
expression for the moments in a pure Coulomb approximation
\be
{\cal M}_n^C=3\pi N_c\left({\Gamma\left(n-\frac{1}{2}\right)\over
4\Gamma\left(\frac{1}{2}\right)\Gamma\left(n+1\right)}
+{C_F\al_s\over n}
+\sum_{m=2}^\infty \left({C_F\al_s\over 2}\right)^m
{\zeta(m)
\Gamma\left(n+\frac{m-1}{2}\right)\over
\Gamma\left(\frac{m-1}{2}\right)\Gamma\left(n+1\right)}\right).
\label{pureCoul}
\ee
The terms of the series in eq.~(\ref{pureCoul}) first increase
numerically 
and then decrease. The number $\tilde m$
of the term with a maximal numerical magnitude is
determined from the relation 
\be
\label{max}
\Psi_1 \left(n+\frac{\tilde m-1}{2}\right)-\Psi_1\left(
\frac{\tilde m-1}{2}\right)
=2\ln\left({2\over C_F\al_s}\right).
\ee
In eq.~(\ref{max}) 
we neglect a weak $m$ dependence of $\zeta(m)$ for large $m$.   
For several first terms with $m<n$ and large $n$ the 
coefficients of the series increase by $\sqrt{n}$ in subsequent orders of the 
expansion that makes the Coulomb resummation necessary. 
Note that for large $m>n$ and fixed $n$ 
the character of asymptotic behavior in $m$ changes and terms become
proportional to
\be
\left({C_F\al_s\over 2}\right)^m \zeta(m)
\left(\frac{m-1}{2}\right)^n
\label{pCas}
\ee
and the  series~(\ref{pureCoul}) is convergent in $\al_s$
at $C_F\al_s/2<1$.

On the other hand, no resummation is necessary for small $n$
and finite order perturbation theory is sufficient for the 
phenomenological analysis giving a reasonable precision.
Moreover, the nonrelativistic approximation is not relevant
for the analysis of low moments.
The ordinary perturbation theory  
expressions for several first moments of the spectral density
are now available with $\al_s^2$ accuracy \cite{Chet}. 
These low moments, however, cannot be used in theoretical formulas for 
sum rules directly because they get a
sizable contribution from the large momentum region far from threshold.  
In this case, the experimental moment necessary for comparison 
can not be found with sufficiently high precision 
because the spectrum is not well known experimentally.

   For numerical estimates it is important to fix 
the allowed range for the normalization point  
which is present in the explicit formula of the polarization
function. Following the general line of 
the renormalization group approach 
the  normalization point has to be chosen
to minimize the higher order corrections.
In fact, the normalization points of $\al_s$
entering the coefficient $C_h$ and the nonrelativistic Green function
can be different when NNLO corrections are considered. The difference
between the normalization points of the hard and soft  
corrections can be noticed only in higher orders 
of perturbative expansion. At first 
sight this gives an additional possibility to improve the convergence
of the perturbation theory for the moments. The typical hard scale
of the problem is the heavy quark mass $m_b$. 
Indeed, one can see that for $\mu\sim m_b$
the NNLO  correction to $C_h$ is  small
in comparison with the  NLO one. 
However a  naive estimate of the soft  ``physical scale''
as a characteristic scale of the 
Coulomb problem $\mu\sim m_b\al_s$ \cite{Vol,H} 
is not acceptable since the direct calculation of the NNLO corrections
shows that the perturbation theory series for the moments blow 
up for this value of the soft normalization point in 
$\overline{\rm MS}$ renormalization scheme.  
This phenomenon can be clearly seen from Fig.~1 and Fig.~2
where the relative weights of the NLO and NNLO corrections
to the parameters of the nonrelativistic Green function  
are plotted. Though the NLO corrections reach their minimal magnitude  
at $\mu\sim m_b\al_s$ the NNLO corrections to the Green function
and the moments are completely out of control at this point.
At first glance this seems to contradict our physical intuition.
However since the normalization scale is defined in the rather 
artificial $\overline {\rm MS}$ scheme the connection of which with
$b\bar b$ physics is not straightforward 
there is no reason for coincidence of $\mu$
parameter with any physical scale of the process.
The relative weight of the 
NNLO corrections is stabilized at $\mu\sim m_b$. Moreover,
here the $\mu$ dependence of the moments is minimal   
so in our opinion there is no reason to split the 
hard and soft normalization points and use the smaller numerical value
for the
soft normalization scale. The dependence of the moments on 
$\mu_f$ is rather weak and we put $\mu_f=\mu$.

The  range of $n$ which can be used  for reliable 
estimates is also quite restricted.   
Indeed, the low moments cannot
be used in sum rules because of the large uncertainty on the
experimental side due to the poor knowledge of the spectral density at
large energies as has been already pointed out.  
From Table~\ref{tab1} 
one sees that for $n$\raisebox{-3pt}{$\stackrel{<}{\sim}$}$8$ 
the experimental moments are rather sensitive 
to the form of the continuos spectrum beyond the resonance region. 
Note that we assume rather large uncertainty of the continuum 
to be on the safe side. However, its contribution is essentially
suppressed in comparison with the resonance one 
and the resulting error of the whole quantity in eq.~(\ref{expmom})
is of the same order of magnitude
as the uncertainties introduced by the resonance contribution.

\begin{table}[htb]
\begin{center}
\begin{tabular}{|c|c|c|c|} \hline
$t$  & ${\cal M}_5^{exp}$ & ${\cal M}_{10}^{exp}$ & 
${\cal M}_{15}^{exp}$ \\ \hline
0.5        & 1.185  & 1.015  & 1.057  \\
1.0        & 1.315  & 1.028  & 1.059  \\
1.5        & 1.446  & 1.041  & 1.061  \\ \hline
\end{tabular}
\end{center}
\caption{Sensitivity  of the experimental moments
to the continuum contribution above $s_0$ for $s_0=(11.2~{\rm GeV})^2$.} 
\label{tab1}
\end{table}

On the other hand, the leading 
nonperturbative power correction to the polarization function 
due to gluonic condensate which can introduce a
large uncertainty on the theoretical side of sum rules
is known to be important for $n>20$
\cite{NSVZ}. The smallness of the nonperturbative 
contribution happened to be the only practical restriction
on maximal allowed $n$ in NLO analysis \cite{Vol}.
In NNLO analysis, however, the upper limit on $n$
is stronger and is connected with the behavior
of the perturbative expansion for the moments as well.    
From Tables~\ref{tab2},~\ref{tab3} one sees that
for $n$\raisebox{-3pt}{$\stackrel{>}{\sim}$}$12$ 
the perturbative series for the moments 
is not well convergent and the $\mu$ dependence
of the moments becomes strong that can be considered 
as an indication that the higher order corrections are large
here. Perhaps this can  mean that the Coulomb solution
is not a good leading order approximation 
for these $n$. 
A possible explanation of this phenomenon can be
obtained from the analysis of an 
asymptotic character of the series in $\al_s$ for the moments.
Indeed, the Coulomb resummation~(\ref{pureCoul}) extracts from
the coefficients of perturbation theory for the moments
the part that forms a convergent
series. This means that resummation
cannot change the analytic structure of the moment
as a function of the coupling constant.  On the other hand  
the Coulomb approximation for the $n$th moment  
is saturated by the large  terms
of the order $m\sim \tilde m$  
where  $\tilde m$ is the solution of 
eq.~(\ref{max}). Now suppose that the full perturbative expansion
for the moments is an asymptotic series with 
$m_{opt}$ being the critical order where the series 
starts to diverge. In this case if  the number of the moment 
is large enough
{\it i.e.} if   $\tilde m$ is close to  $m_{opt}$
the asymptotic growth of the high order coefficients of the full series
becomes more important than the resummed Coulomb part  of the series.    
For such $n$ the Coulomb approximation  does not saturate the 
full series coefficients in the orders
which are dominant in the Coulomb approximation itself
and large corrections naturally appear.     
The condition for applicability
of the Coulomb solution as a leading order approximation now
is $\tilde m<m_{opt}$. Taking into account that 
the perturbative expansion around the Coulomb solution
is not well convergent  for $n$\raisebox{-3pt}{$\stackrel{>}{\sim}$}$12$ 
from the above condition and eq.~(\ref{max})
for the physical value of $\al_s$ in $\overline {\rm MS}$ scheme
one finds $m_{opt}\sim 3-4$ which is a reasonable value
for a QCD series. 

\begin{table}[htb]
\begin{center}
\begin{tabular}{|l|c|c|c|c|} \hline
{Approx.} & ${\cal M}_5^{th}$ & ${\cal M}_{10}^{th}$ & 
${\cal M}_{15}^{th}$ & ${\cal M}_{20}^{th}$  \\ \hline
LO         & 0.9319  & 0.6679  & 0.5602 & 0.5169 \\
NLO        & 0.7905  & 0.6633  & 0.6413 & 0.6682 \\
NLO$^*$    & 0.8123  & 0.7235  & 0.7590 & 0.8694 \\
NNLO       & 1.047   & 0.973   & 1.026  & 1.150  \\
NNLO$^*$   & 1.147   & 1.297   & 1.760  & 2.601  \\ \hline
\end{tabular}
\end{center}
\caption{ The 0th, 1st and 2nd order theoretical  moments for
$\al_s(M_z)=0.118$, $m_b=4.8~{\rm GeV}$ and $\mu=m_b$.
The continuum contribution above $s_0$ is subtracted. Here the star $(^*)$
stands for the moments obtained by using  the  
representation~(\ref{endenom}) for Green function.} 
\label{tab2}
\end{table}

\begin{table}[htb]
\begin{center}
\begin{tabular}{|c|c|c|c|c|} \hline
$\mu$ (GeV) & ${\cal M}_5^{th}$ & ${\cal M}_{10}^{th}$ & 
${\cal M}_{15}^{th}$ & ${\cal M}_{20}^{th}$        \\ \hline
6        & 0.870  & 0.764  & 0.769  & 0.827\\
$m_b$    & 1.047  & 0.973  & 1.026  & 1.150  \\
4        & 1.259  & 1.220  & 1.334  & 1.543 \\
3        & 1.761  & 1.867  & 2.209  & 2.735 \\ \hline
\end{tabular}
\end{center}
\caption{The scale dependence of the  theoretical  moments
in the NNLO approximation for $\al_s(M_z)=0.118$ and 
$m_b=4.8~{\rm GeV}$ (the continuum contribution  above $s_0$
is subtracted).} 
\label{tab3}
\end{table}

Here a remark concerning a specific form of Green function is in order.  
The difference between two representations (\ref{totcorr}) and
(\ref{endenom}) for the Green function appears 
only in higher orders in the coupling constant.   
We do not attemt to compute the spectral density
at positive energies 
in the threshold region because the nonperturbative 
effects are large and prevent point-wise evaluation of this quantity. 
We are interested in the polarization 
function in Euclidean region where some high order derivatives can be
obtained. 
The nonrelativistic Green function
has no immediate meaning for us as a source for 
the spectral density at positive energies in threshold region 
and only serves as a tool 
for resummation of a special contributions 
of perturbation theory expansion. Therefore these two representations
are equivalent in NNLO. We find, however, that 
when the representation~(\ref{endenom}) is used the perturbative
series for the moments are more divergent especially for 
large $n$ (see Table~\ref{tab2}) and the dependence 
of the results on $n$ and $\mu$ is stronger.
Therefore we consider the representation~(\ref{totcorr}) 
as a preferable one. The fact that  resummation of  the
corrections keeping the Coulomb form of the nonrelativistic Green function
spoils the properties of the perturbative expansion
can be considered as another indication that for  
large $n$ the Coulomb resummation is not enough and 
a more adequate leading order approximation is necessary with 
a proper large $n$ behavior.   

Now we turn to a discussion of numerical values of parameters
extracted from the analysis of sum rules.  
The sum rules for the $\Upsilon$ system are not very sensitive to $\al_s$ 
so NLO \cite{KPP} and NNLO \cite{PP} analyses  give a rather rough 
estimate
\be
\al_s(M_Z)=0.118 \pm 0.006
\label{alfin}
\ee
which is in agreement with other available data \cite{PDG}.
This result is obtained by
simultaneous fit for $\al_s$ and $m_b$. The central values
are found by the standard least $\chi^2$ method \cite{Vol}. 
The uncertainty of the fit is estimated  
by fixing $m_b$ to its central value and varying 
$\mu$ and $n$ within allowed range.  
On the other hand, the sum rules are much more sensitive to the
$b$-quark mass  so it is instructive
to fix $\al_s$ to the ``world average'' value 
$\al_s(M_Z)=0.118$ \cite{PDG}
and then extract $m_b$.  
The final estimate of the bottom quark pole mass 
is\footnote{In ref.~\cite{PP} the  interval 
$ 4.74~{\rm GeV}\le m_b\le 4.82~{\rm GeV}$ has been obtained
for $b$-quark pole mass. In the present paper we use the 
different range of $n$ and $\mu$ for the analysis of the sum rules 
and give more conservative estimate of the uncertainty.}  
\be
m_b=4.80\pm 0.06 ~{\rm GeV} .
\label{mbfin}
\ee 
The uncertainty corresponds to the interval 
$3.5~{\rm GeV}\le\mu\le 6.5~{\rm GeV}$ for $8\le n\le 12$. 
In fact the central value changes slightly in a wider 
interval $5<n<20$ but we restrict the range of  $n$ 
to minimize   the uncertainty  related to $\mu$ and $t$
dependence. Note that the scale dependence of the moments   
is mainly due to the scale   dependence of $\al_s(\mu)$ 
while the explicit dependence  on $\mu$ is rather
weak. 

The perturbation theory 
relation between the perturbative pole mass and the
mass $\overline{m}_b(\mu)$ defined in the $\overline{\rm MS}$ renormalization scheme is 
known up to 
two-loop level \cite{GBGS}.
It is therefore straightforward to
find the numerical value for the $\overline{\rm MS}$ 
mass of $b$-quark using result of eq.~(\ref{mbfin}) 
\[
\overline{m}_b(\overline{m}_b)=4.21\pm 0.11 ~{\rm GeV} 
\]
where we assume the interval~(\ref{alfin})
for the strong coupling constant. An interesting fact is that
the convergence of the perturbative series for 
$\overline{\rm MS}$ mass obtained order by order from the 
$\Upsilon$ sum rules
\[
\overline{m}_b(\overline{m}_b)
=(\overline{m}_b(\overline{m}_b))^{LO}(1-0.085-0.021)
\]
is much better than for the pole mass
\[
m_b=m_b^{LO}(1-0.001+0.021) 
\]
though the second order terms in the above expansions are rather 
close numerically.   

Let us emphasize 
that the convergence of the perturbation theory
for  the vacuum polarization function of heavy quark
near the threshold and to the moments is not fast
in the $\overline{\rm MS}$ renormalization scheme.
We have found the NNLO
corrections to exceed the NLO ones.
Furthermore, in the case of $b$-quark the corrections  due to the 
perturbative modification of the Coulomb 
instantaneous potential ({\it i.e.} related to $\Delta G_1$ and 
$\Delta G_2^{(i)}$ terms) dominate
the total correction in the NLO and 
NNLO. Inclusion of these corrections is quite important for 
consistent analysis of sum rules for the $\Upsilon$
system. 
A conjecture that this fact is a consequence of the 
asymptotic character of the series which leads to the intrinsic
ambiguity in the heavy quark pole mass was studied in detail in
the literature \cite{MY,ren,JPS,ren2}.
It mainly based on consideration of the renormalon contribution  
\cite{ren,ren2}. 
Within this picture the high order contributions
to the moments are saturated by the corrections that are associated
with $(\al_s\beta_0\ln(\mu r))^m$ terms in perturbative
series for the heavy quark potential. These corrections are generated 
by terms stemming from the running of the coupling constant that can
be found from renormalization group analysis. 
However, the explicit 
calculation shows that in NNLO 
the corresponding contribution (the one proportional to the 
coefficient $C_2^2$) provides only $\sim 10\%$ of the total
NNLO correction {\it i.e.} the NNLO correction turns out to be
essentially larger than the renormalon picture predicts.  
Thus the asymptotic series  for the pole mass  seems to 
reach its critical order and starts to diverge.
If this is a case then eq.~(\ref{mbfin}) gives a numerical
estimate of a finite order  sum of the asymptotic series. 
This value extracted from the sum rules
can be used as an auxiliary parameter in the expressions for 
the physical quantities order by order of QCD perturbation theory.
This, however, is sufficient for physical
applications because the pole mass is not an observable.
 
An independent NNLO analysis of 
$\Upsilon$ sum rules has been done in ref.~\cite{H}
where the Laplace transform of the polarization function
was studied in the spirit of ref.~\cite{Vol} while the explicit
NNLO expression of the polarization function near the threshold
has not been obtained. We found that for a given set of 
the parameters the numerical values of the theoretical
moments obtained on the basis of our result for 
the polarization function is in a good agreement
with the results of ref.~\cite{H}. However, in ref.~\cite{H}
an  essentially lower range of the soft normalization
scale $1.5~{\rm GeV}<\mu <3.5~{\rm GeV}$ was used
for the phenomenological analysis. 
This does not affect 
strongly the result for the strong coupling constant
$0.96<\al_s(M_Z)<0.124$ obtained in ref.~\cite{H} but leads to 
larger value of the $b$-quark pole mass  
$4.78~{\rm GeV}<m_b<4.98~{\rm GeV}$ 
(the result of the constrained fit for the fixed $\al_s$ of ref.~\cite{H}) 
than the number given by eq.~(\ref{mbfin}).
From Fig.~2 we, however, see that such a low normalization 
scale is unsuitable for a reliable estimate because
the perturbation theory around the Coulomb solution
diverges in this case. 

\section{Inclusive $B$-meson semileptonic width and  $|V_{cb}|$
matrix element.}
\label{inclusive} 
 
The theoretical expression for the inclusive
semileptonic width of $B$-meson
up to the second order in the strong coupling constant and up to the first
order of heavy quark expansion reads
\be
\Gamma_{\rm sl}={G_F^2m_b^5\over 192\pi^3}|V_{cb}|^2\left(F_1\left(
{m_c^2/m_b^2}\right)C_\Gamma(\alpha_s)
\left(1-{\mu_\pi^2-\mu_G^2\over 2m_b^2}\right)-
2F_2\left({m_c^2/m_b^2}\right)
{\mu_G^2\over m_b^2}\right)
\label{Gamma}
\ee 
where $F_1(x)=1-8x-12x^2\ln{x}+8x^3-x^4$ is the phase space factor, 
$F_2(x)=(1-x)^4$,
$\mu_\pi$ and  
$\mu_G$  are the  
HQET parameters \cite{np,MW}. 
The hadronic matrix element of 
the gluon dipole operator $\mu_G$
of the heavy quark 
is directly
related to the masses of $B$-mesons with different spin structure
\[
\mu_G^2\left(1+O(1/m_b)\right)={4\over 3}(M^2_{B^*}-M^2_{B})=0.36~{\rm GeV^2}.
\] 
The hadronic matrix element of the heavy quark 
kinetic operator $\mu_\pi$ suffers from larger uncertainty. 
For numerical estimates we use the result 
\[
\mu_\pi^2= (0.5\pm 0.15)~{\rm GeV^2} 
\] 
obtained within QCD sum rules framework \cite{BBBG}.
The perturbative coefficient $C_\Gamma(\alpha_s)$ up to the second
order in $\al_s$ \cite{CM,Nir} is
\be
C_\Gamma=1-1.67{\al_s(\mu)\over\pi}
-(8.4\pm 0.4)\left({\al_s\over\pi}\right)^2
\label{cg}
\ee
for the normalization point
$\mu=\sqrt{m_bm_c}$. 

Along with ordinary power corrections coming from 
heavy quark expansion an account for truncation 
of perturbative series for semileptonic width 
can generate further contributions to eq.~(\ref{Gamma})
which within the renormalon picture scale as $\Lambda_{QCD}/m_b$.   
We discuss some details related to this issue later.

One sees 
from formula~(\ref{Gamma})  
that the semileptonic width depends
rather strongly on $m_b$.
Therefore if $m_b$ is taken from a theoretical expression 
for some other experimental quantity it should
be determined with a great accuracy for obtaining
a reasonable precision of
the $|V_{cb}|$ determination. In the previous Section 
we have described the moments of the vacuum polarization function and
have extracted the numerical value of the $b$-quark pole mass
entering the theoretical perturbation theory expressions
for the moments.
Being analyzed independently, 
the perturbative series in $\al_s$ for 
the moments and for the width expressed in terms of the pole mass $m_b$
seem not to enjoy a fast apparent
convergence that can lead to a large uncertainty due to 
higher order contributions.
In principle the apparent convergence of the finite order 
perturbation theory series can be changed
by redefinition of the mass \cite{ren,lsm,BSU} or the coupling constant 
\cite{KPP,U1}.
However $m_b$ is not an observable and has no
immediate physical meaning. Therefore it can be safely removed from 
relations between physical observables. The idea of trading the
unphysical parameters in favor of direct relations between
observables constitutes now the
most general trend in high precision phenomenological analyses,
its application to the
particular case of the $B$-meson semileptonic decay width 
was discussed earlier (see {\it e.g.}~\cite{U,BSU}
and references therein).  
In this Section we establish the direct relation between 
inclusive $B$-meson semileptonic width
and one of the moments obtained in the previous section.
This relation is independent of any redefinition of the 
quark mass and of using different couplings 
so the accuracy of the relation reveals the actual precision of the
approximation used for comparison of two physical observables. 
It happens that the convergence of the perturbation theory
for this relation is rather fast and the subsequent
approximations converge very well in a heuristic sence that
the higher order approximation is close to the previous one.

Our analysis consists in 
direct relating the factor $m_b^5$ in eq.~(\ref{Gamma}) to
the $n$th moment of the $\Upsilon$ sum rules 
\be
m_b^5
=\left({{\cal M}^{th}_n\over \tilde{{\cal M}}_n^{exp}}\right)^{5\over 2n}.
\label{mb}
\ee 
where (dimensionful) moments 
$\tilde{{\cal M}}_n^{exp}={\cal M}_n^{exp}/m_b^{2n}$ is a purely experimental
quantity. 
The theoretical moment ${\cal M}^{th}_n$ is a dimensionless quantity 
which depends on $m_b$  only 
logarithmically
in a finite order in $\al_s$.  
Thus, the substitution of relation~(\ref{mb}) to eq.~(\ref{Gamma})
substantially reduces its dependence on $m_b$ though it also introduces 
an explicit uncertainty due to ${\cal M}^{exp}_n$. 
 
We use eqs.~(\ref{Gamma},\ref{mb}) to find for the  mixing angle 
$|V_{cb}|$ 
\be
|V_{cb}| 
=({192 \pi^3})^{1/2}K_{th}
{\G_{\rm sl}^{1/2}({\tilde M^{exp}_n})^{5/4n}\over G_F}
\label{vcb}
\ee
where the functions
\be
K_{th}
=({M^{th}_n})^{-5/4n}
\left(F_1\left(
{m_c^2/m_b^2}\right)C_\Gamma(\alpha_s)
\left(1-{\mu_\pi^2-\mu_G^2\over 2m^2_b}\right)-
2F_2\left({m_c^2/m_b^2}\right)
{\mu_G^2\over m_b^2}\right)^{-{1\over 2}}
\label{Kcoef}
\ee
accumulate
theoretical information depending on
$m_b$, $m_c$, $\alpha_s$, etc.

The function $F_1(m_c^2/m_b^2)$ gives
rather large theoretical uncertainty if masses of $b$- and
$c$-quarks are considered as independent variables.
However there is almost model independent constraint of the form
\be
m_b-m_c-\mu_\pi^2\left({1\over 2m_c}-{1\over 2m_b}\right)+O(1/m^2_{b,c})=
\bar M_B - \bar M_D=3.34~{\rm GeV}
\label{const}
\ee
where $\bar M_B=5.31~{\rm GeV}$, $\bar M_D=1.97~{\rm GeV}$ denote
the spin-average meson masses, e.g. $\bar M_B=\frac{1}{4}(M_B+
3M_{B^*})$. 
With this constraint the function $F_1(m_c^2/m_b^2)$ 
becomes a function of a single variable
$\tilde{F}_1(m_b)$. 
Note that in such a setting the $m_b$ dependence of 
the function $\tilde F_1(m_b)$ 
partly cancels the large $m^5_b$ dependence of the width.
Furthermore if the relation ~(\ref{mb})
is used to express the $b$-quark pole mass in the argument of the function 
$\tilde F_1(m_b)$ in terms of the moments of the spectral 
density then only logarithmic  
dependence on $m_b$ appears in the right hand side of eq.~(\ref{vcb}).

Now we analyze the theoretical factor 
$K_{th}$
numerically order by order in $\al_s$. The result reads
\be
K_{th}|_{n=10}
=1.366(1+0.088+0.028) .
\label{thseries}
\ee
where the value of $b$-quark pole
mass in  the fixed order in $\al_s$ with Coulomb resummation
is found from   
eq.~(\ref{mb}). For comparison, the perturbative series
for $m_b$ that follows from eq.~(\ref{mb}) and the series 
for $C_\Gamma$ are
\be
m_b|_{n=10}=4.71(1-0.001+0.021) ,
\label{mbseries}
\ee
\be
C_\Gamma=1-0.146-0.064 .
\label{cgseries}
\ee
Thus we find that in the  
expansions of the theoretical moment (as well as $m_b$ itself) 
and width expressed in terms of $m_b$ the NNLO corrections
are of the order of the NLO ones
while   the perturbative series for the mixing angle,
or the theoretical coefficient eq.~(\ref{thseries}),
converges much better. 

Let us discuss the problem of convergence in more details. 
Though there is no rigorous result on the asymptotic structure
of the expansion~(\ref{cg}) it is widely believed 
(mainly due to consideration of the renormalon contribution) 
that this asymptotic series (as well as the series for the  moments) 
starts to diverge already in some low orders of perturbation theory. 
The truncation of the asymptotic series at the optimal order
results in the inherent uncertainty in these quantities
which parametrically as large as  
$\Lambda_{QCD}/m_b$ 
within the renormalon 
picture. 
An argument in favor of this conjecture is the relatively
large value of the NNLO corrections both to the 
moments and the semileptonic width.
Since the detailed structure of the above
asymptotic expansions remains unknown we cannot conclude that
the divergence of the series for the moments and the series
for the width cancels each other exactly. But  because 
the terms in the perturbative expansion~(\ref{thseries}) 
decrease rapidly we can reasonably hope that
the critical order where the series~(\ref{thseries}) 
starts to diverge is higher than ones of eq.~(\ref{mbseries}) and
eq.~(\ref{cgseries}). Moreover, a partial cancellation 
of the divergences has been found within the renormalon 
picture \cite{MY,ren2}.  

As for numerics, we use the following central values 
for our experimental inputs (see \cite{PDG} for more detail):
\[
{\rm BR}(B\rightarrow X_cl\nu_l)=10.5\%, \qquad \tau_B = 1.55~{\rm ps},
\]
\[
({M^{exp}_{10}})^{1/4}=3.95\times 10^{-4}~{\rm GeV}^{-5}, \qquad
\al_s(M_Z)=0.118 .
\]
With these numbers we obtain
the value of the matrix element $|V_{cb}|$
\be
|V_{cb}|=0.0423\left({{\rm BR}(B\rightarrow X_cl\nu_l)
\over 0.105}\right)^{1\over 2}
\left({1.55 {\rm ps}\over \tau_B}\right)^{1\over 2}
\label{final}
\ee
\[
\times\left(1 -0.01{\al_s(M_Z)-0.118 \over 0.006}\right)
\left(1\pm\Delta_{npt}\right)\left(1\pm\Delta_{tr}\right)
\]
where $\Delta_{npt}\sim 0.02$
is the uncertainty in the nonperturbative contribution \cite{U} 
induced mainly by the uncertainty of $\mu_\pi$
in eq.~(\ref{const}).
We have introduced also the uncertainty due to truncation
of the perturbative series for the $K_{th}$ parameter 
$\Delta_{tr}\sim 0.01$ which is taken as
a half of the last term in eq.~(\ref{thseries}).

The typical scale of uncertainty of key parameters is also
indicated.
Another important source 
of the uncertainty is 
the scale dependence of the theoretical moment 
and the experimental errors 
in the value of $\al_s(M_Z)$ because of rather high
sensitivity of the theoretical moment to $\al_s$.
In fact these uncertainties are closely related since
as it has already been pointed out  
the scale dependence of ${\cal M}_{10}^{th}(\al_s(\mu),\mu)$ 
is mainly due to the scale   dependence of $\al_s(\mu)$. 
The pointed error bars roughly correspond to the interval 
$3.5~{\rm GeV}\le\mu\le 6.5~{\rm GeV}$  at 
fixed $\al_s(M_Z)=0.118$.
The central value in eq.~(\ref{final}) does not change 
in the interval $5\le n\le 15$ and we have chosen $n=10$
for our final estimate by the reasons 
discussed in the previous section.    

The main part of the experimental uncertainty is 
related to the uncertainty in the experimentally measured 
inclusive semileptonic width.
The experimental situation changes rather quickly 
and data are improving fast
that means that the experimental uncertainties will be smaller 
(see e.g. \cite{prosp}).
The uncertainty in $M^{exp}_{10}$ comes mainly from 
the uncertainties in leptonic widths $\Gamma_{k}$ of $\Upsilon$ 
resonances. It is about $5\%$
and leads only to $0.3\%$ uncertainty in $|V_{cb}|$ so we do not 
include it to the error bars in eq.~(\ref{final}).

Now the advantage of our approach becomes clear --
large errors due to the uncertainty in $m_b$ 
is now partly shifted to more direct experimental data.
The expression for $|V_{cb}|$ matrix element 
has a very weak dependence on $m_b$ so
a possible uncertainty in $b$-quark pole mass 
does not
lead to any uncertainty in $|V_{cb}|$.
Furthermore the  terms in the perturbative expansion~(\ref{thseries}) 
decrease rapidly which is an indication 
that the  higher  orders corrections 
to the obtained result are small enough.   

Our result is in a good agreement 
with the previous estimate $|V_{cb}|=0.0419$ \cite{U}. 
Our value, however,
is somewhat larger than the estimate $|V_{cb}|=0.039$ 
of ref.~\cite{N}.  
There is no much hope to reduce the uncertainty in the 
nonperturbative contribution. Thus, in our opinion, the model 
independent result presented in the paper
provides one with the most reliable and accurate estimate of 
the CKM matrix element $|V_{cb}|$ from
the inclusive $B$-meson semileptonic width.

\section{Conclusion.}
\label{secconcl}
In this paper we have presented the determination of the $b$-quark pole and 
$\overline{\rm MS}$ mass and 
Cabibbo-Kobayashi-Maskawa matrix element $|V_{cb}|$
from  $\Upsilon$ sum rules and $B$-meson semileptonic width in the 
NNLO in the strong coupling constant.
In our opinion it seems to be a final result for these problems
within the framework of analytical treatment:
the next order approximation is too
complicated to deal with analytically within QCD
or NRQCD. In particular, the nonrelativistic approximation in next order 
in $\al_s$ should be supplemented by the real gluon radiation 
that takes the problem to the completely new level of complexity 
and makes it practically unsolvable within 
the framework presented here. 

The self-consistence of the  
$\Upsilon$ sum rules has been checked and the intervals
for the relevant normalization scale and the moment numbers which can
be used for a reliable estimates have been found. 
An asymptotic character of the perturbative expansion for the moments 
of the $\Upsilon$ system spectral density was discussed and a
conjecture on the critical order where the series in $\al_s$ 
for the moments starts to diverge has been made. 
We have also presented a new representation
of the heavy quark vector current correlator
near threshold based on the explicit formulas for the NNLO correction
to the $^3S_1$ heavy quark bound state parameters\footnote{When this 
paper was in preparation a paper \cite{MY} appeared where the similar 
result was obtained.}. 

We have constructed the direct relation
between the moments of the $\Upsilon$ system spectral
density and the inclusive $B$-meson semileptonic width up to the  
NNLO order. We have found that when the unphysical variable
(the $b$-quark pole mass) is removed 
the residual perturbation theory in the coupling
constant for $\Gamma_{\rm sl}$ (or $|V_{cb}|$) 
works well and demonstrates nice convergence.

\vspace{7mm}
\noindent
{\large \bf Acknowledgments}\\[2mm]
We thank J.H.K{\"u}hn
for support, encouragement, and discussions.
A.A.Penin gratefully acknowledges discussions
with K.Melnikov. This work is partially supported 
by Volkswagen Foundation under contract
No.~I/73611. A.A.Pivo\-varov is  supported in part by
the Russian Fund for Basic Research under contracts Nos.~96-01-01860
and 97-02-17065. The work of A.A.Penin is supported in part  by
the Russian Fund for Basic Research under contract
97-02-17065.

\newpage
\section*{Appendix.}
{\large\bf A.} The correction $\Delta_2^{(2)}G$  due to the $\Delta_2V$ 
part of the potential is (see eq.~(\ref{totcorr}))
\[
\Delta_2^{(2)}G=\left({\al_s\over 4\pi}\right)^2{C_F\al_sm_b^2\over
4\pi}\left( \sum_{m=0}^\infty F^2(m)
\left((m+1)\left(C_0^2+L(k) C_1^2+L^2(k) C_2^2\right)
\right.\right.
\]
$$
\left.
+(m+1)\Psi_1(m+2)\left(C_1^2+2L(k)
C_2^2\right)+I(m)C_2^2\right) 
$$
\[
+2\sum_{m=1}^\infty\sum_{n=0}^{m-1}
F(m)F(n)\left(-
{n+1\over m-n}\left(C_1^2 +2L(k) C_2^2\right)
+J(m,n)C_2^2\right)
\]
\[
+2\sum_{m=0}^\infty F(m)
\left(C_0^2+L(k) C_1^2+
(L^2(k)+K(m))C_2^2-(2\gamma_E+
\Psi_1(m+1))
\left(C_1^2+2L(k) C_2^2\right)\right)
\]
$$
\left.+L(k)C_0^2+\left(-\gamma_E L(k)+
{1\over 2}L^2(k)\right)
C_1^2 
+N(k)C_2^2\right)
$$
where
\[
\int e^{-z}L^1_n(z)L^1_m(z)\ln^2(z)zdz=
\left\{
\begin{array}{c}
I(m), \quad m=n, \\
J(m,n), \quad m>n, \\
\end{array} 
\right. 
\]
\[
\int e^{-z}L^1_n(z)\ln(z)dz=K(m)
\]
\[
\int {e^{-2kx}\ln^2(\mu x)\over x}dx=N(k)+\ldots
\]
where dots stand for inessential $k$ independent divergent part
and
\[
I(m)=(m+1)\left(\Psi^2_1(m+2)-\Psi_2(m+2)+{\pi^2\over3}-{2\over(m+1)^2}\right)
\]
$$
-2(\Psi_1(m+1)+\gamma_E),
$$
$$
J(m,n)= 2{n+1\over m-n}\left(\Psi_1(m-n)-{1\over n+1}+2\gamma_E\right)
$$
\[
+2{m+1\over m-n}(\Psi_1(m-n+1)-\Psi_1(m+1)),
\]
$$
K(m)=2(\Psi_1(m+1)+\gamma_E)^2+\Psi_2(m+1)-\Psi_1^2(m+1)+2\gamma_E^2,
$$
$$
N(k)=\left(\gamma_E+{\pi^2\over 6}\right)L(k)
-\gamma_E L^2(k)+{1\over 3}
L^3(k).
$$
The correction $\Delta_2^{(1)}G$ due to the 
second iteration of $\Delta_1V$ term 
\[
\Delta_2^{(1)}G=\left({\al_s\over 4\pi}\right)^2
{(C_F\al_s)^2\over 4\pi}{m_b^3\over 2k}
\left( \sum_{m=0}^\infty H^3(m)(m+1)
\right.
\]
$$
\left(C_0^1+
\left(\Psi(m+2)+
L(k)\right)C_1^1\right)^2
$$
\[
-2\sum_{m=1}^\infty\sum_{n=0}^{m-1}{n+1\over m-n}C_1^1
\left(H^2(m)H(n)\left(C_0^1+\left(\Psi(m+2)+
L(k)-{1\over 2}{1\over m-n}\right)C_1^1\right)
\right.
\]
\[
\left.
+H(m)H^2(n)\left(C_0^1+\left(\Psi(n+2)+
L(k)-{1\over 2}{n+1\over (m-n)(m+1)}\right)C_1^1\right)\right)
\]
\[
+2(C_1^1)^2\left(\sum_{m=2}^\infty\sum_{l=1}^{m-1}\sum_{n=0}^{l-1}
{H(m)H(n)H(l)}{n+1\over (l-n) (m-n)}\right.
\]
$$
+\sum_{m=2}^\infty\sum_{n=1}^{m-1}\sum_{l=0}^{n-1}
{H(m)H(n)H(l)}{l+1\over (n-l)(m-n)}
$$
\[
\left.\left.
+\sum_{n=2}^\infty\sum_{m=1}^{n-1}\sum_{l=0}^{m-1}
{H(m)H(n)H(l)}{(l+1)(m+1)\over (n+1)(n-l)(n-m)}\right)\right)
\]

\vspace{5mm}
\noindent
{\large\bf B.} The NNLO corrections to the square 
of the  Coulomb  $^3S_1$ heavy quark 
bound state wave function
at the origin  (eq.~(\ref{ecorr}))
\[
\Delta_{d5}\psi^2_m={1\over 3}{C_F^2\al_s^2\over(m+1)^2},
\]
\[
\Delta_{{\bf \Delta}^2,NA,BF}\psi^2_m=-{C_F^2\al_s^2}
\left(
{15\over 8}{1\over(m+1)^2}+\left({2\over 3}+{C_A\over C_F}\right)
\left(-\ln\left({2\mu_{f}(m+1)\over C_F\al_sm_b}\right)\right.\right.
\]
\[
\left.\left.
+\gamma_E
+\Psi_1(m+1)-{1\over(m+1)}\right)
\right),
\]
\[
\Delta^{(2)}_2\psi^2_m=\left({\alpha_s\over 4\pi}\right)^2
\left(3(C^2_0+L(m)C_1^2+L^2(m)C^2_2) +
(-1-2\gamma_E+{2\over m+1}+\Psi_1(m+2)\right.
\]
\[
-2(m+1)\Psi_2(m+1))(C_1^2+2L(m)C^2_2)+\left({I(m)\over m+1}+2K(m)-2\Psi_1(m+2)\right.
\]
\[
\left.\left.
+2\sum_{n=0}^{m-1}{m+1\over (n-m)(n+1)}J(m,n)+2\sum_{n=m+1}^\infty{m+1\over (n-m)(n+1)}J(n,m)
\right)C_2^2
\right),
\]
\[
\Delta^{(1)}_2\psi^2_m=\left({\alpha_s\over 4\pi}\right)^2
\left(3(C_0^1+(L(m)+
\Psi_1(m+2))C_1^1)^2\right.
\]
\[
+2C_1^1\left(\sum_{n=0}^{m-1}{(n+1)(m+1)\over (n-m)^3}\left(
C_0^1+\left((L(m)+
\Psi_1(n+2)+{1\over 2}{n+1\over (n-m)(m+1)}\right)C_1^1\right)
\right.
\]
\[
\left.
-\sum_{n=m+1}^{\infty}{(m+1)^2\over (n-m)^3}\left(
C_0^1+\left((L(m)+
\Psi_1(n+2)-{1\over 2}{1\over n-m}\right)C_1^1\right)
\right)
\]
\[
+\left.
2C_1^1\left(C_0^1+\left(L(m)+\Psi_1(m+2)\right)
C_1^1\right)
\left(-{5\over 2}+\sum_{n=0}^{m-1}{n+1\over (n-m)^2}U(m,n)
\right.
\right.
\]
\[
\left.
-\sum_{n=m+1}^{\infty}{m+1\over (n-m)^2}U(m,n)
\right)
+2(C_1^1)^2\left({1\over 2}-\sum_{n=0}^{m-1}{n+1\over (n-m)^2}
+\sum_{n=m+1}^{\infty}{m+1\over (n-m)^2}\right.
\]
\[
\left.
+{1\over 2}\sum_{n=0}^{m-1}{n+1\over (n-m)^3}U(m,n)
+{1\over 2}\sum_{n=m+1}^{\infty}{(m+1)^2\over (n-m)^3(n+1)}U(m,n)
\right.
\]
\[
+\sum_{n=1}^{m-1}\sum_{l=0}^{n-1}\left(
{(l+1)(n+1)\over (n-m)^2(l-m)^2}-{(l+1)(m+1)\over (n-m)^2(l-m)(n-l)}-
{(l+1)(m+1)\over (l-m)^2(n-m)(n-l)}\right)
\]
\[
+\sum_{n=m+1}^{\infty}\sum_{l=0}^{m-1}\left(
-{(l+1)(m+1)\over (n-m)^2(l-m)^2}+{(l+1)(m+1)^2\over (n-m)^2(l-m)(n-l)(n+1)}
\right.
\]
\[
\left.
-{(l+1)(m+1)\over (l-m)^2(n-m)(n-l)}\right)
+\sum_{n=2}^{\infty}\sum_{l=m+1}^{n-1}\left(
{(m+1)^2\over (n-m)^2(l-m)^2}\right.
\]
\[
\left.\left.\left.
+{(l+1)(m+1)^2\over (n-m)^2(l-m)(n-l)(n+1)}+
{(m+1)^2\over (l-m)^2(n-m)(n-l)}\right)\right)\right)
\]
where
\[
U(m,n)=3+{n+1\over m+n+2}
-2{(n+1)^2\over (n-m)(n+m+2)}
\]

\vspace{5mm}
\noindent
{\large\bf C.} The NNLO corrections to the   Coulomb  $^3S_1$
heavy quark
bound state energy levels (eq.~(\ref{pscorr}))
\[
\Delta_{{\bf \Delta}^2,NA,BF}E_m={C_F^2\al_s^2\over (m+1)}
\left({C_A\over C_F}+{2\over 3} -{11\over 16}{1\over (m+1)}\right),
\]
\[
\Delta_2^{(2)}E_m=2\left({\alpha_s\over 4\pi}\right)^2
\left(C_0^2+L(m)C_1^2+L^2(m)C_2^2+
\Psi_1(m+2)(C_1^2+2L(m)C_2^2)\right.
\]
\[
\left.
+{I(m)\over (m+1)}C_2^2\right),
\]
\[
\Delta_2^{(1)}E_m=\left({\alpha_s\over 4\pi}\right)^2
\left(\left(C_0^1+(L(m)-2+
\Psi_1(m+2))C_1^1\right)
\left(C_0^1+(L(m)
+\Psi_1(m+2))C_1^1\right)\right.
\]
\[
\left.
+\left({2\over  (m+1)}(\gamma_E+\Psi_1(m+2))
-2\Psi_2(m+1)-(m+1)\Psi_3(m+1)\right)(C_1^1)^2\right),
\]
The results for the heavy quark bound state
parameters are obtained on the basis of corrections 
found in \cite{PP}.
When this work was in preparation ref.~\cite{MY} appeared
where the correction to the  $^1S_3$ heavy quark  bound state
wave function at the origin were computed up to NNLO.
This calculation is consistent with our result for 
$\Delta_{1}\psi^2_m$,
$\Delta_{d5}\psi^2_m$,  $\Delta_{NA,BF,\Delta}\psi^2_m$
and  $\Delta^{(2)}_2\psi^2_m$. The expression for   $\Delta^{(1)}_2\psi^2_m$
was obtained in ref.~\cite{MY}  in a different representation
so direct comparison with our formula is difficult.
At the same time numerically the results are in a good agreement.  
Our result for the NNLO correction to the heavy quark $^1S_3$ bound state
energy levels coincides with the result of ref.~\cite{PY}.
 
\section*{Figure captions}

\noindent
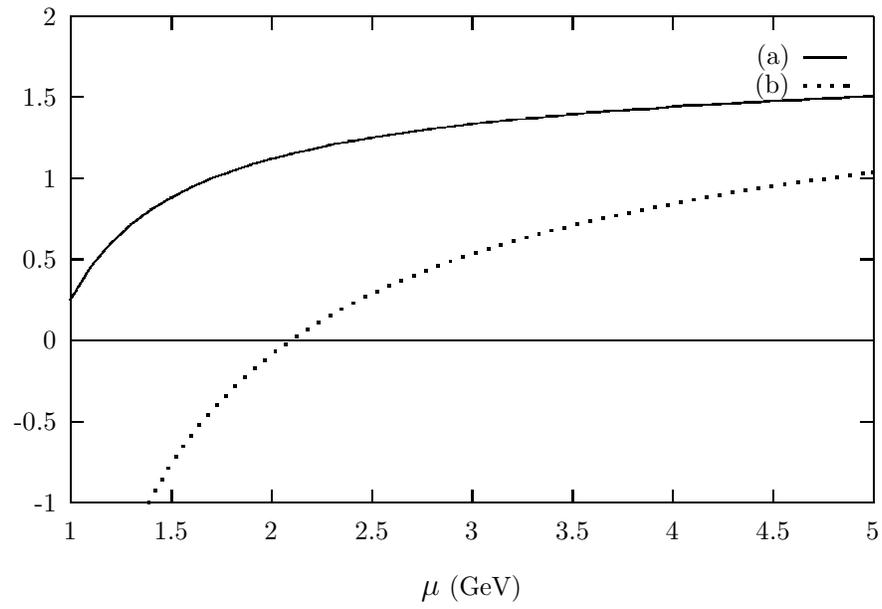
{\bf Fig. 1.}
The relative weight of the 
NLO corrections to the ground state energy 
$E_0^{NLO}/E_0^{LO}-1$  (curve {\it a})  
and to the square of the ground state  wave function
at the origin\\   
$(|\psi_0(0)|^2)^{NLO}/(|\psi_0(0)|^2)^{LO}-1$ 
(curve {\it b}) as a function of the  normalization 
point $\mu$.

\noindent
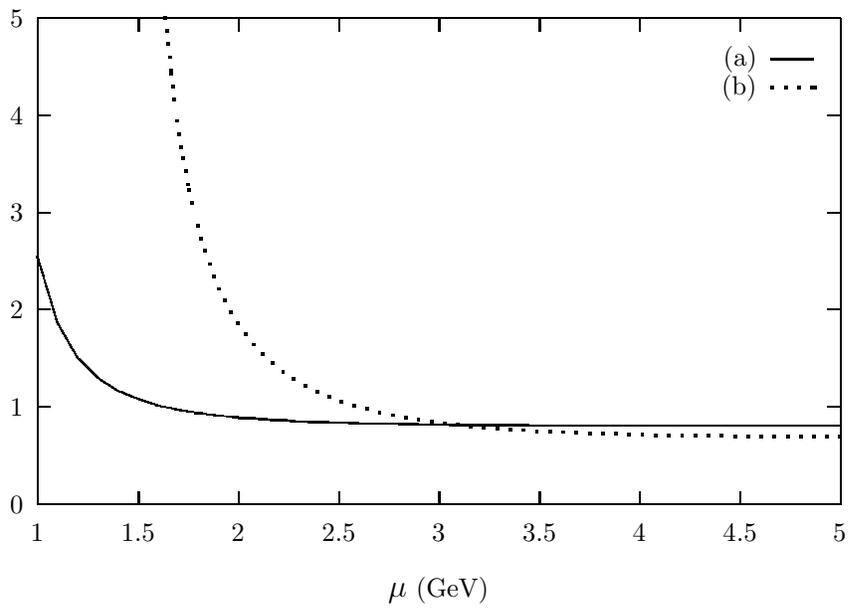
{\bf Fig. 2.}
The relative weight of the 
NNLO corrections to the ground state energy\\ 
$E_0^{NNLO}/E_0^{NLO}-1$  (curve {\it a}) and 
to the square of the ground state  wave function
at the origin   
$(|\psi_0(0)|^2)^{NNLO}/(|\psi_0(0)|^2)^{NLO}-1$ 
(curve {\it b}) as a function of the  normalization 
point $\mu$.

\newpage
\begin{center}

\setlength{\unitlength}{0.240900pt}
\ifx\plotpoint\undefined\newsavebox{\plotpoint}\fi
\sbox{\plotpoint}{\rule[-0.200pt]{0.400pt}{0.400pt}}%
\begin{picture}(1500,900)(0,0)
\font\gnuplot=cmr10 at 10pt
\gnuplot
\sbox{\plotpoint}{\rule[-0.200pt]{0.400pt}{0.400pt}}%
\put(176.0,368.0){\rule[-0.200pt]{303.534pt}{0.400pt}}
\put(176.0,113.0){\rule[-0.200pt]{4.818pt}{0.400pt}}
\put(154,113){\makebox(0,0)[r]{-1}}
\put(1416.0,113.0){\rule[-0.200pt]{4.818pt}{0.400pt}}
\put(176.0,240.0){\rule[-0.200pt]{4.818pt}{0.400pt}}
\put(154,240){\makebox(0,0)[r]{-0.5}}
\put(1416.0,240.0){\rule[-0.200pt]{4.818pt}{0.400pt}}
\put(176.0,368.0){\rule[-0.200pt]{4.818pt}{0.400pt}}
\put(154,368){\makebox(0,0)[r]{0}}
\put(1416.0,368.0){\rule[-0.200pt]{4.818pt}{0.400pt}}
\put(176.0,495.0){\rule[-0.200pt]{4.818pt}{0.400pt}}
\put(154,495){\makebox(0,0)[r]{0.5}}
\put(1416.0,495.0){\rule[-0.200pt]{4.818pt}{0.400pt}}
\put(176.0,622.0){\rule[-0.200pt]{4.818pt}{0.400pt}}
\put(154,622){\makebox(0,0)[r]{1}}
\put(1416.0,622.0){\rule[-0.200pt]{4.818pt}{0.400pt}}
\put(176.0,750.0){\rule[-0.200pt]{4.818pt}{0.400pt}}
\put(154,750){\makebox(0,0)[r]{1.5}}
\put(1416.0,750.0){\rule[-0.200pt]{4.818pt}{0.400pt}}
\put(176.0,877.0){\rule[-0.200pt]{4.818pt}{0.400pt}}
\put(154,877){\makebox(0,0)[r]{2}}
\put(1416.0,877.0){\rule[-0.200pt]{4.818pt}{0.400pt}}
\put(176.0,113.0){\rule[-0.200pt]{0.400pt}{4.818pt}}
\put(176,68){\makebox(0,0){1}}
\put(176.0,857.0){\rule[-0.200pt]{0.400pt}{4.818pt}}
\put(334.0,113.0){\rule[-0.200pt]{0.400pt}{4.818pt}}
\put(334,68){\makebox(0,0){1.5}}
\put(334.0,857.0){\rule[-0.200pt]{0.400pt}{4.818pt}}
\put(491.0,113.0){\rule[-0.200pt]{0.400pt}{4.818pt}}
\put(491,68){\makebox(0,0){2}}
\put(491.0,857.0){\rule[-0.200pt]{0.400pt}{4.818pt}}
\put(649.0,113.0){\rule[-0.200pt]{0.400pt}{4.818pt}}
\put(649,68){\makebox(0,0){2.5}}
\put(649.0,857.0){\rule[-0.200pt]{0.400pt}{4.818pt}}
\put(806.0,113.0){\rule[-0.200pt]{0.400pt}{4.818pt}}
\put(806,68){\makebox(0,0){3}}
\put(806.0,857.0){\rule[-0.200pt]{0.400pt}{4.818pt}}
\put(964.0,113.0){\rule[-0.200pt]{0.400pt}{4.818pt}}
\put(964,68){\makebox(0,0){3.5}}
\put(964.0,857.0){\rule[-0.200pt]{0.400pt}{4.818pt}}
\put(1121.0,113.0){\rule[-0.200pt]{0.400pt}{4.818pt}}
\put(1121,68){\makebox(0,0){4}}
\put(1121.0,857.0){\rule[-0.200pt]{0.400pt}{4.818pt}}
\put(1279.0,113.0){\rule[-0.200pt]{0.400pt}{4.818pt}}
\put(1279,68){\makebox(0,0){4.5}}
\put(1279.0,857.0){\rule[-0.200pt]{0.400pt}{4.818pt}}
\put(1436.0,113.0){\rule[-0.200pt]{0.400pt}{4.818pt}}
\put(1436,68){\makebox(0,0){5}}
\put(1436.0,857.0){\rule[-0.200pt]{0.400pt}{4.818pt}}
\put(176.0,113.0){\rule[-0.200pt]{303.534pt}{0.400pt}}
\put(1436.0,113.0){\rule[-0.200pt]{0.400pt}{184.048pt}}
\put(176.0,877.0){\rule[-0.200pt]{303.534pt}{0.400pt}}
\put(806,-22){\makebox(0,0){$\mu$ (GeV)}}
\put(176.0,113.0){\rule[-0.200pt]{0.400pt}{184.048pt}}
\put(1306,812){\makebox(0,0)[r]{(a)}}
\put(1328.0,812.0){\rule[-0.200pt]{15.899pt}{0.400pt}}
\put(176,431){\usebox{\plotpoint}}
\multiput(176.58,431.00)(0.497,0.815){61}{\rule{0.120pt}{0.750pt}}
\multiput(175.17,431.00)(32.000,50.443){2}{\rule{0.400pt}{0.375pt}}
\multiput(208.58,483.00)(0.497,0.597){59}{\rule{0.120pt}{0.577pt}}
\multiput(207.17,483.00)(31.000,35.802){2}{\rule{0.400pt}{0.289pt}}
\multiput(239.00,520.58)(0.532,0.497){57}{\rule{0.527pt}{0.120pt}}
\multiput(239.00,519.17)(30.907,30.000){2}{\rule{0.263pt}{0.400pt}}
\multiput(271.00,550.58)(0.675,0.496){43}{\rule{0.639pt}{0.120pt}}
\multiput(271.00,549.17)(29.673,23.000){2}{\rule{0.320pt}{0.400pt}}
\multiput(302.00,573.58)(0.847,0.495){35}{\rule{0.774pt}{0.119pt}}
\multiput(302.00,572.17)(30.394,19.000){2}{\rule{0.387pt}{0.400pt}}
\multiput(334.00,592.58)(0.977,0.494){29}{\rule{0.875pt}{0.119pt}}
\multiput(334.00,591.17)(29.184,16.000){2}{\rule{0.438pt}{0.400pt}}
\multiput(365.00,608.58)(1.121,0.494){25}{\rule{0.986pt}{0.119pt}}
\multiput(365.00,607.17)(28.954,14.000){2}{\rule{0.493pt}{0.400pt}}
\multiput(396.00,622.58)(1.486,0.492){19}{\rule{1.264pt}{0.118pt}}
\multiput(396.00,621.17)(29.377,11.000){2}{\rule{0.632pt}{0.400pt}}
\multiput(428.00,633.58)(1.439,0.492){19}{\rule{1.227pt}{0.118pt}}
\multiput(428.00,632.17)(28.453,11.000){2}{\rule{0.614pt}{0.400pt}}
\multiput(459.00,644.59)(1.834,0.489){15}{\rule{1.522pt}{0.118pt}}
\multiput(459.00,643.17)(28.841,9.000){2}{\rule{0.761pt}{0.400pt}}
\multiput(491.00,653.59)(2.079,0.488){13}{\rule{1.700pt}{0.117pt}}
\multiput(491.00,652.17)(28.472,8.000){2}{\rule{0.850pt}{0.400pt}}
\multiput(523.00,661.59)(2.323,0.485){11}{\rule{1.871pt}{0.117pt}}
\multiput(523.00,660.17)(27.116,7.000){2}{\rule{0.936pt}{0.400pt}}
\multiput(554.00,668.59)(2.323,0.485){11}{\rule{1.871pt}{0.117pt}}
\multiput(554.00,667.17)(27.116,7.000){2}{\rule{0.936pt}{0.400pt}}
\multiput(585.00,675.59)(2.841,0.482){9}{\rule{2.233pt}{0.116pt}}
\multiput(585.00,674.17)(27.365,6.000){2}{\rule{1.117pt}{0.400pt}}
\multiput(617.00,681.59)(3.493,0.477){7}{\rule{2.660pt}{0.115pt}}
\multiput(617.00,680.17)(26.479,5.000){2}{\rule{1.330pt}{0.400pt}}
\multiput(649.00,686.59)(3.382,0.477){7}{\rule{2.580pt}{0.115pt}}
\multiput(649.00,685.17)(25.645,5.000){2}{\rule{1.290pt}{0.400pt}}
\multiput(680.00,691.59)(3.493,0.477){7}{\rule{2.660pt}{0.115pt}}
\multiput(680.00,690.17)(26.479,5.000){2}{\rule{1.330pt}{0.400pt}}
\multiput(712.00,696.60)(4.429,0.468){5}{\rule{3.200pt}{0.113pt}}
\multiput(712.00,695.17)(24.358,4.000){2}{\rule{1.600pt}{0.400pt}}
\multiput(743.00,700.60)(4.429,0.468){5}{\rule{3.200pt}{0.113pt}}
\multiput(743.00,699.17)(24.358,4.000){2}{\rule{1.600pt}{0.400pt}}
\multiput(774.00,704.60)(4.575,0.468){5}{\rule{3.300pt}{0.113pt}}
\multiput(774.00,703.17)(25.151,4.000){2}{\rule{1.650pt}{0.400pt}}
\multiput(806.00,708.61)(6.937,0.447){3}{\rule{4.367pt}{0.108pt}}
\multiput(806.00,707.17)(22.937,3.000){2}{\rule{2.183pt}{0.400pt}}
\multiput(838.00,711.61)(6.714,0.447){3}{\rule{4.233pt}{0.108pt}}
\multiput(838.00,710.17)(22.214,3.000){2}{\rule{2.117pt}{0.400pt}}
\multiput(869.00,714.61)(6.714,0.447){3}{\rule{4.233pt}{0.108pt}}
\multiput(869.00,713.17)(22.214,3.000){2}{\rule{2.117pt}{0.400pt}}
\multiput(900.00,717.61)(6.937,0.447){3}{\rule{4.367pt}{0.108pt}}
\multiput(900.00,716.17)(22.937,3.000){2}{\rule{2.183pt}{0.400pt}}
\multiput(932.00,720.61)(6.937,0.447){3}{\rule{4.367pt}{0.108pt}}
\multiput(932.00,719.17)(22.937,3.000){2}{\rule{2.183pt}{0.400pt}}
\multiput(964.00,723.61)(6.714,0.447){3}{\rule{4.233pt}{0.108pt}}
\multiput(964.00,722.17)(22.214,3.000){2}{\rule{2.117pt}{0.400pt}}
\put(995,726.17){\rule{6.500pt}{0.400pt}}
\multiput(995.00,725.17)(18.509,2.000){2}{\rule{3.250pt}{0.400pt}}
\put(1027,728.17){\rule{6.300pt}{0.400pt}}
\multiput(1027.00,727.17)(17.924,2.000){2}{\rule{3.150pt}{0.400pt}}
\put(1058,730.17){\rule{6.300pt}{0.400pt}}
\multiput(1058.00,729.17)(17.924,2.000){2}{\rule{3.150pt}{0.400pt}}
\multiput(1089.00,732.61)(6.937,0.447){3}{\rule{4.367pt}{0.108pt}}
\multiput(1089.00,731.17)(22.937,3.000){2}{\rule{2.183pt}{0.400pt}}
\put(1121,735.17){\rule{6.300pt}{0.400pt}}
\multiput(1121.00,734.17)(17.924,2.000){2}{\rule{3.150pt}{0.400pt}}
\put(1152,736.67){\rule{7.709pt}{0.400pt}}
\multiput(1152.00,736.17)(16.000,1.000){2}{\rule{3.854pt}{0.400pt}}
\put(1184,738.17){\rule{6.300pt}{0.400pt}}
\multiput(1184.00,737.17)(17.924,2.000){2}{\rule{3.150pt}{0.400pt}}
\put(1215,740.17){\rule{6.500pt}{0.400pt}}
\multiput(1215.00,739.17)(18.509,2.000){2}{\rule{3.250pt}{0.400pt}}
\put(1247,742.17){\rule{6.500pt}{0.400pt}}
\multiput(1247.00,741.17)(18.509,2.000){2}{\rule{3.250pt}{0.400pt}}
\put(1279,743.67){\rule{7.468pt}{0.400pt}}
\multiput(1279.00,743.17)(15.500,1.000){2}{\rule{3.734pt}{0.400pt}}
\put(1310,745.17){\rule{6.500pt}{0.400pt}}
\multiput(1310.00,744.17)(18.509,2.000){2}{\rule{3.250pt}{0.400pt}}
\put(1342,746.67){\rule{7.468pt}{0.400pt}}
\multiput(1342.00,746.17)(15.500,1.000){2}{\rule{3.734pt}{0.400pt}}
\put(1373,748.17){\rule{6.500pt}{0.400pt}}
\multiput(1373.00,747.17)(18.509,2.000){2}{\rule{3.250pt}{0.400pt}}
\put(1405,749.67){\rule{7.468pt}{0.400pt}}
\multiput(1405.00,749.17)(15.500,1.000){2}{\rule{3.734pt}{0.400pt}}
\sbox{\plotpoint}{\rule[-0.500pt]{1.000pt}{1.000pt}}%
\put(1306,767){\makebox(0,0)[r]{(b)}}
\multiput(1328,767)(20.756,0.000){4}{\usebox{\plotpoint}}
\put(1394,767){\usebox{\plotpoint}}
\put(297.00,113.00){\usebox{\plotpoint}}
\put(307.02,131.16){\usebox{\plotpoint}}
\put(317.47,149.09){\usebox{\plotpoint}}
\multiput(318,150)(10.679,17.798){0}{\usebox{\plotpoint}}
\put(328.23,166.84){\usebox{\plotpoint}}
\put(339.67,184.15){\usebox{\plotpoint}}
\put(351.52,201.17){\usebox{\plotpoint}}
\put(363.63,218.02){\usebox{\plotpoint}}
\multiput(365,220)(12.453,16.604){0}{\usebox{\plotpoint}}
\put(376.25,234.48){\usebox{\plotpoint}}
\put(389.85,250.15){\usebox{\plotpoint}}
\multiput(393,254)(13.962,15.358){0}{\usebox{\plotpoint}}
\put(403.61,265.68){\usebox{\plotpoint}}
\put(417.81,280.81){\usebox{\plotpoint}}
\multiput(422,285)(14.676,14.676){0}{\usebox{\plotpoint}}
\put(432.56,295.41){\usebox{\plotpoint}}
\put(447.65,309.65){\usebox{\plotpoint}}
\multiput(450,312)(15.513,13.789){0}{\usebox{\plotpoint}}
\put(463.21,323.37){\usebox{\plotpoint}}
\multiput(469,328)(15.513,13.789){0}{\usebox{\plotpoint}}
\put(479.06,336.74){\usebox{\plotpoint}}
\put(495.36,349.54){\usebox{\plotpoint}}
\multiput(497,351)(17.004,11.902){0}{\usebox{\plotpoint}}
\put(512.29,361.52){\usebox{\plotpoint}}
\multiput(516,364)(17.004,11.902){0}{\usebox{\plotpoint}}
\put(529.40,373.27){\usebox{\plotpoint}}
\multiput(535,377)(17.004,11.902){0}{\usebox{\plotpoint}}
\put(546.51,385.01){\usebox{\plotpoint}}
\multiput(554,390)(18.144,10.080){0}{\usebox{\plotpoint}}
\put(564.25,395.75){\usebox{\plotpoint}}
\put(581.78,406.85){\usebox{\plotpoint}}
\multiput(582,407)(18.564,9.282){0}{\usebox{\plotpoint}}
\put(600.14,416.52){\usebox{\plotpoint}}
\multiput(601,417)(18.564,9.282){0}{\usebox{\plotpoint}}
\put(618.51,426.17){\usebox{\plotpoint}}
\multiput(620,427)(18.564,9.282){0}{\usebox{\plotpoint}}
\put(636.88,435.82){\usebox{\plotpoint}}
\multiput(639,437)(19.271,7.708){0}{\usebox{\plotpoint}}
\put(655.61,444.67){\usebox{\plotpoint}}
\multiput(658,446)(18.967,8.430){0}{\usebox{\plotpoint}}
\put(674.59,453.04){\usebox{\plotpoint}}
\multiput(677,454)(18.967,8.430){0}{\usebox{\plotpoint}}
\put(693.72,461.09){\usebox{\plotpoint}}
\multiput(696,462)(18.967,8.430){0}{\usebox{\plotpoint}}
\put(712.84,469.14){\usebox{\plotpoint}}
\multiput(715,470)(18.967,8.430){0}{\usebox{\plotpoint}}
\put(731.97,477.19){\usebox{\plotpoint}}
\multiput(734,478)(19.690,6.563){0}{\usebox{\plotpoint}}
\put(751.30,484.69){\usebox{\plotpoint}}
\multiput(752,485)(19.880,5.964){0}{\usebox{\plotpoint}}
\put(770.72,491.88){\usebox{\plotpoint}}
\multiput(771,492)(19.880,5.964){0}{\usebox{\plotpoint}}
\multiput(781,495)(19.690,6.563){0}{\usebox{\plotpoint}}
\put(790.49,498.20){\usebox{\plotpoint}}
\multiput(800,502)(19.690,6.563){0}{\usebox{\plotpoint}}
\put(809.98,505.29){\usebox{\plotpoint}}
\multiput(819,508)(19.690,6.563){0}{\usebox{\plotpoint}}
\put(829.77,511.53){\usebox{\plotpoint}}
\multiput(838,514)(19.690,6.563){0}{\usebox{\plotpoint}}
\put(849.54,517.85){\usebox{\plotpoint}}
\multiput(856,520)(20.352,4.070){0}{\usebox{\plotpoint}}
\put(869.56,523.19){\usebox{\plotpoint}}
\multiput(875,525)(19.880,5.964){0}{\usebox{\plotpoint}}
\put(889.47,528.99){\usebox{\plotpoint}}
\multiput(894,530)(19.880,5.964){0}{\usebox{\plotpoint}}
\put(909.38,534.79){\usebox{\plotpoint}}
\multiput(913,536)(20.352,4.070){0}{\usebox{\plotpoint}}
\put(929.40,540.13){\usebox{\plotpoint}}
\multiput(932,541)(20.261,4.503){0}{\usebox{\plotpoint}}
\put(949.62,544.72){\usebox{\plotpoint}}
\multiput(951,545)(19.690,6.563){0}{\usebox{\plotpoint}}
\put(969.67,549.93){\usebox{\plotpoint}}
\multiput(970,550)(20.261,4.503){0}{\usebox{\plotpoint}}
\multiput(979,552)(19.880,5.964){0}{\usebox{\plotpoint}}
\put(989.75,555.17){\usebox{\plotpoint}}
\multiput(998,557)(20.352,4.070){0}{\usebox{\plotpoint}}
\put(1010.05,559.46){\usebox{\plotpoint}}
\multiput(1017,561)(20.261,4.503){0}{\usebox{\plotpoint}}
\put(1030.33,563.87){\usebox{\plotpoint}}
\multiput(1036,565)(20.261,4.503){0}{\usebox{\plotpoint}}
\put(1050.64,568.13){\usebox{\plotpoint}}
\multiput(1055,569)(20.261,4.503){0}{\usebox{\plotpoint}}
\put(1070.96,572.39){\usebox{\plotpoint}}
\multiput(1074,573)(20.261,4.503){0}{\usebox{\plotpoint}}
\put(1091.27,576.65){\usebox{\plotpoint}}
\multiput(1093,577)(20.261,4.503){0}{\usebox{\plotpoint}}
\put(1111.58,580.92){\usebox{\plotpoint}}
\multiput(1112,581)(20.261,4.503){0}{\usebox{\plotpoint}}
\multiput(1121,583)(20.629,2.292){0}{\usebox{\plotpoint}}
\put(1132.01,584.40){\usebox{\plotpoint}}
\multiput(1140,586)(20.261,4.503){0}{\usebox{\plotpoint}}
\put(1152.32,588.66){\usebox{\plotpoint}}
\multiput(1159,590)(20.629,2.292){0}{\usebox{\plotpoint}}
\put(1172.80,591.96){\usebox{\plotpoint}}
\multiput(1178,593)(20.261,4.503){0}{\usebox{\plotpoint}}
\put(1193.20,595.62){\usebox{\plotpoint}}
\multiput(1197,596)(20.261,4.503){0}{\usebox{\plotpoint}}
\put(1213.53,599.67){\usebox{\plotpoint}}
\multiput(1215,600)(20.652,2.065){0}{\usebox{\plotpoint}}
\put(1233.98,603.00){\usebox{\plotpoint}}
\multiput(1234,603)(20.652,2.065){0}{\usebox{\plotpoint}}
\multiput(1244,604)(20.261,4.503){0}{\usebox{\plotpoint}}
\put(1254.46,606.15){\usebox{\plotpoint}}
\multiput(1263,607)(20.261,4.503){0}{\usebox{\plotpoint}}
\put(1274.94,609.29){\usebox{\plotpoint}}
\multiput(1282,610)(20.261,4.503){0}{\usebox{\plotpoint}}
\put(1295.42,612.44){\usebox{\plotpoint}}
\multiput(1301,613)(20.261,4.503){0}{\usebox{\plotpoint}}
\put(1315.89,615.65){\usebox{\plotpoint}}
\multiput(1319,616)(20.652,2.065){0}{\usebox{\plotpoint}}
\put(1336.40,618.64){\usebox{\plotpoint}}
\multiput(1338,619)(20.652,2.065){0}{\usebox{\plotpoint}}
\multiput(1348,620)(20.629,2.292){0}{\usebox{\plotpoint}}
\put(1357.01,621.00){\usebox{\plotpoint}}
\multiput(1367,623)(20.629,2.292){0}{\usebox{\plotpoint}}
\put(1377.50,624.15){\usebox{\plotpoint}}
\multiput(1386,625)(20.261,4.503){0}{\usebox{\plotpoint}}
\put(1397.98,627.30){\usebox{\plotpoint}}
\multiput(1405,628)(20.629,2.292){0}{\usebox{\plotpoint}}
\put(1418.62,629.51){\usebox{\plotpoint}}
\multiput(1423,630)(20.352,4.070){0}{\usebox{\plotpoint}}
\put(1433,632){\usebox{\plotpoint}}
\end{picture}

\vspace{5mm}
{\bf Fig. 1.}
\vspace{10mm}

\setlength{\unitlength}{0.240900pt}
\ifx\plotpoint\undefined\newsavebox{\plotpoint}\fi
\sbox{\plotpoint}{\rule[-0.200pt]{0.400pt}{0.400pt}}%
\begin{picture}(1500,900)(0,0)
\font\gnuplot=cmr10 at 10pt
\gnuplot
\sbox{\plotpoint}{\rule[-0.200pt]{0.400pt}{0.400pt}}%
\put(176.0,113.0){\rule[-0.200pt]{303.534pt}{0.400pt}}
\put(176.0,113.0){\rule[-0.200pt]{4.818pt}{0.400pt}}
\put(154,113){\makebox(0,0)[r]{0}}
\put(1416.0,113.0){\rule[-0.200pt]{4.818pt}{0.400pt}}
\put(176.0,266.0){\rule[-0.200pt]{4.818pt}{0.400pt}}
\put(154,266){\makebox(0,0)[r]{1}}
\put(1416.0,266.0){\rule[-0.200pt]{4.818pt}{0.400pt}}
\put(176.0,419.0){\rule[-0.200pt]{4.818pt}{0.400pt}}
\put(154,419){\makebox(0,0)[r]{2}}
\put(1416.0,419.0){\rule[-0.200pt]{4.818pt}{0.400pt}}
\put(176.0,571.0){\rule[-0.200pt]{4.818pt}{0.400pt}}
\put(154,571){\makebox(0,0)[r]{3}}
\put(1416.0,571.0){\rule[-0.200pt]{4.818pt}{0.400pt}}
\put(176.0,724.0){\rule[-0.200pt]{4.818pt}{0.400pt}}
\put(154,724){\makebox(0,0)[r]{4}}
\put(1416.0,724.0){\rule[-0.200pt]{4.818pt}{0.400pt}}
\put(176.0,877.0){\rule[-0.200pt]{4.818pt}{0.400pt}}
\put(154,877){\makebox(0,0)[r]{5}}
\put(1416.0,877.0){\rule[-0.200pt]{4.818pt}{0.400pt}}
\put(176.0,113.0){\rule[-0.200pt]{0.400pt}{4.818pt}}
\put(176,68){\makebox(0,0){1}}
\put(176.0,857.0){\rule[-0.200pt]{0.400pt}{4.818pt}}
\put(334.0,113.0){\rule[-0.200pt]{0.400pt}{4.818pt}}
\put(334,68){\makebox(0,0){1.5}}
\put(334.0,857.0){\rule[-0.200pt]{0.400pt}{4.818pt}}
\put(491.0,113.0){\rule[-0.200pt]{0.400pt}{4.818pt}}
\put(491,68){\makebox(0,0){2}}
\put(491.0,857.0){\rule[-0.200pt]{0.400pt}{4.818pt}}
\put(649.0,113.0){\rule[-0.200pt]{0.400pt}{4.818pt}}
\put(649,68){\makebox(0,0){2.5}}
\put(649.0,857.0){\rule[-0.200pt]{0.400pt}{4.818pt}}
\put(806.0,113.0){\rule[-0.200pt]{0.400pt}{4.818pt}}
\put(806,68){\makebox(0,0){3}}
\put(806.0,857.0){\rule[-0.200pt]{0.400pt}{4.818pt}}
\put(964.0,113.0){\rule[-0.200pt]{0.400pt}{4.818pt}}
\put(964,68){\makebox(0,0){3.5}}
\put(964.0,857.0){\rule[-0.200pt]{0.400pt}{4.818pt}}
\put(1121.0,113.0){\rule[-0.200pt]{0.400pt}{4.818pt}}
\put(1121,68){\makebox(0,0){4}}
\put(1121.0,857.0){\rule[-0.200pt]{0.400pt}{4.818pt}}
\put(1279.0,113.0){\rule[-0.200pt]{0.400pt}{4.818pt}}
\put(1279,68){\makebox(0,0){4.5}}
\put(1279.0,857.0){\rule[-0.200pt]{0.400pt}{4.818pt}}
\put(1436.0,113.0){\rule[-0.200pt]{0.400pt}{4.818pt}}
\put(1436,68){\makebox(0,0){5}}
\put(1436.0,857.0){\rule[-0.200pt]{0.400pt}{4.818pt}}
\put(176.0,113.0){\rule[-0.200pt]{303.534pt}{0.400pt}}
\put(1436.0,113.0){\rule[-0.200pt]{0.400pt}{184.048pt}}
\put(176.0,877.0){\rule[-0.200pt]{303.534pt}{0.400pt}}
\put(806,-22){\makebox(0,0){$\mu$ (GeV)}}
\put(176.0,113.0){\rule[-0.200pt]{0.400pt}{184.048pt}}
\put(1306,812){\makebox(0,0)[r]{(a)}}
\put(1328.0,812.0){\rule[-0.200pt]{15.899pt}{0.400pt}}
\put(176,502){\usebox{\plotpoint}}
\multiput(176.58,496.14)(0.497,-1.652){61}{\rule{0.120pt}{1.413pt}}
\multiput(175.17,499.07)(32.000,-102.068){2}{\rule{0.400pt}{0.706pt}}
\multiput(208.58,393.69)(0.497,-0.874){59}{\rule{0.120pt}{0.797pt}}
\multiput(207.17,395.35)(31.000,-52.346){2}{\rule{0.400pt}{0.398pt}}
\multiput(239.00,341.92)(0.499,-0.497){61}{\rule{0.500pt}{0.120pt}}
\multiput(239.00,342.17)(30.962,-32.000){2}{\rule{0.250pt}{0.400pt}}
\multiput(271.00,309.92)(0.778,-0.496){37}{\rule{0.720pt}{0.119pt}}
\multiput(271.00,310.17)(29.506,-20.000){2}{\rule{0.360pt}{0.400pt}}
\multiput(302.00,289.92)(1.250,-0.493){23}{\rule{1.085pt}{0.119pt}}
\multiput(302.00,290.17)(29.749,-13.000){2}{\rule{0.542pt}{0.400pt}}
\multiput(334.00,276.92)(1.590,-0.491){17}{\rule{1.340pt}{0.118pt}}
\multiput(334.00,277.17)(28.219,-10.000){2}{\rule{0.670pt}{0.400pt}}
\multiput(365.00,266.93)(2.323,-0.485){11}{\rule{1.871pt}{0.117pt}}
\multiput(365.00,267.17)(27.116,-7.000){2}{\rule{0.936pt}{0.400pt}}
\multiput(396.00,259.93)(3.493,-0.477){7}{\rule{2.660pt}{0.115pt}}
\multiput(396.00,260.17)(26.479,-5.000){2}{\rule{1.330pt}{0.400pt}}
\multiput(428.00,254.94)(4.429,-0.468){5}{\rule{3.200pt}{0.113pt}}
\multiput(428.00,255.17)(24.358,-4.000){2}{\rule{1.600pt}{0.400pt}}
\multiput(459.00,250.95)(6.937,-0.447){3}{\rule{4.367pt}{0.108pt}}
\multiput(459.00,251.17)(22.937,-3.000){2}{\rule{2.183pt}{0.400pt}}
\put(491,247.17){\rule{6.500pt}{0.400pt}}
\multiput(491.00,248.17)(18.509,-2.000){2}{\rule{3.250pt}{0.400pt}}
\put(523,245.17){\rule{6.300pt}{0.400pt}}
\multiput(523.00,246.17)(17.924,-2.000){2}{\rule{3.150pt}{0.400pt}}
\put(554,243.17){\rule{6.300pt}{0.400pt}}
\multiput(554.00,244.17)(17.924,-2.000){2}{\rule{3.150pt}{0.400pt}}
\put(585,241.67){\rule{7.709pt}{0.400pt}}
\multiput(585.00,242.17)(16.000,-1.000){2}{\rule{3.854pt}{0.400pt}}
\put(617,240.67){\rule{7.709pt}{0.400pt}}
\multiput(617.00,241.17)(16.000,-1.000){2}{\rule{3.854pt}{0.400pt}}
\put(649,239.67){\rule{7.468pt}{0.400pt}}
\multiput(649.00,240.17)(15.500,-1.000){2}{\rule{3.734pt}{0.400pt}}
\put(680,238.67){\rule{7.709pt}{0.400pt}}
\multiput(680.00,239.17)(16.000,-1.000){2}{\rule{3.854pt}{0.400pt}}
\put(743,237.67){\rule{7.468pt}{0.400pt}}
\multiput(743.00,238.17)(15.500,-1.000){2}{\rule{3.734pt}{0.400pt}}
\put(712.0,239.0){\rule[-0.200pt]{7.468pt}{0.400pt}}
\put(806,236.67){\rule{7.709pt}{0.400pt}}
\multiput(806.00,237.17)(16.000,-1.000){2}{\rule{3.854pt}{0.400pt}}
\put(774.0,238.0){\rule[-0.200pt]{7.709pt}{0.400pt}}
\put(932,235.67){\rule{7.709pt}{0.400pt}}
\multiput(932.00,236.17)(16.000,-1.000){2}{\rule{3.854pt}{0.400pt}}
\put(838.0,237.0){\rule[-0.200pt]{22.645pt}{0.400pt}}
\put(964.0,236.0){\rule[-0.200pt]{113.705pt}{0.400pt}}
\sbox{\plotpoint}{\rule[-0.500pt]{1.000pt}{1.000pt}}%
\put(1306,767){\makebox(0,0)[r]{(b)}}
\multiput(1328,767)(20.756,0.000){4}{\usebox{\plotpoint}}
\put(1394,767){\usebox{\plotpoint}}
\multiput(375,877)(2.112,-20.648){5}{\usebox{\plotpoint}}
\multiput(384,789)(2.506,-20.604){3}{\usebox{\plotpoint}}
\multiput(393,715)(3.468,-20.464){3}{\usebox{\plotpoint}}
\multiput(403,656)(3.677,-20.427){3}{\usebox{\plotpoint}}
\multiput(412,606)(4.918,-20.164){2}{\usebox{\plotpoint}}
\put(426.04,549.27){\usebox{\plotpoint}}
\multiput(431,530)(6.563,-19.690){2}{\usebox{\plotpoint}}
\put(444.51,489.85){\usebox{\plotpoint}}
\put(451.45,470.29){\usebox{\plotpoint}}
\multiput(459,451)(9.282,-18.564){2}{\usebox{\plotpoint}}
\put(477.98,414.04){\usebox{\plotpoint}}
\multiput(478,414)(11.000,-17.601){0}{\usebox{\plotpoint}}
\put(489.00,396.45){\usebox{\plotpoint}}
\put(500.81,379.42){\usebox{\plotpoint}}
\put(514.02,363.42){\usebox{\plotpoint}}
\multiput(516,361)(13.962,-15.358){0}{\usebox{\plotpoint}}
\put(527.96,348.04){\usebox{\plotpoint}}
\put(543.43,334.26){\usebox{\plotpoint}}
\multiput(545,333)(15.513,-13.789){0}{\usebox{\plotpoint}}
\put(559.29,320.88){\usebox{\plotpoint}}
\multiput(563,318)(17.798,-10.679){0}{\usebox{\plotpoint}}
\put(576.66,309.56){\usebox{\plotpoint}}
\multiput(582,306)(17.798,-10.679){0}{\usebox{\plotpoint}}
\put(594.34,298.70){\usebox{\plotpoint}}
\multiput(601,295)(18.564,-9.282){0}{\usebox{\plotpoint}}
\put(612.78,289.21){\usebox{\plotpoint}}
\multiput(620,286)(19.271,-7.708){0}{\usebox{\plotpoint}}
\put(631.91,281.15){\usebox{\plotpoint}}
\multiput(639,278)(19.880,-5.964){0}{\usebox{\plotpoint}}
\put(651.42,274.19){\usebox{\plotpoint}}
\multiput(658,272)(19.690,-6.563){0}{\usebox{\plotpoint}}
\put(671.15,267.75){\usebox{\plotpoint}}
\multiput(677,266)(19.690,-6.563){0}{\usebox{\plotpoint}}
\put(691.06,261.99){\usebox{\plotpoint}}
\multiput(696,261)(19.690,-6.563){0}{\usebox{\plotpoint}}
\put(711.11,256.78){\usebox{\plotpoint}}
\multiput(715,256)(20.261,-4.503){0}{\usebox{\plotpoint}}
\put(731.43,252.51){\usebox{\plotpoint}}
\multiput(734,252)(20.261,-4.503){0}{\usebox{\plotpoint}}
\put(751.86,249.02){\usebox{\plotpoint}}
\multiput(752,249)(20.352,-4.070){0}{\usebox{\plotpoint}}
\multiput(762,247)(20.261,-4.503){0}{\usebox{\plotpoint}}
\put(772.19,244.88){\usebox{\plotpoint}}
\multiput(781,244)(20.629,-2.292){0}{\usebox{\plotpoint}}
\put(792.79,242.44){\usebox{\plotpoint}}
\multiput(800,241)(20.629,-2.292){0}{\usebox{\plotpoint}}
\put(813.32,239.57){\usebox{\plotpoint}}
\multiput(819,239)(20.629,-2.292){0}{\usebox{\plotpoint}}
\put(833.97,237.40){\usebox{\plotpoint}}
\multiput(838,237)(20.629,-2.292){0}{\usebox{\plotpoint}}
\put(854.60,235.16){\usebox{\plotpoint}}
\multiput(856,235)(20.652,-2.065){0}{\usebox{\plotpoint}}
\multiput(866,234)(20.629,-2.292){0}{\usebox{\plotpoint}}
\put(875.24,233.00){\usebox{\plotpoint}}
\multiput(885,233)(20.629,-2.292){0}{\usebox{\plotpoint}}
\put(895.93,231.81){\usebox{\plotpoint}}
\multiput(904,231)(20.629,-2.292){0}{\usebox{\plotpoint}}
\put(916.59,230.00){\usebox{\plotpoint}}
\multiput(923,230)(20.629,-2.292){0}{\usebox{\plotpoint}}
\put(937.29,229.00){\usebox{\plotpoint}}
\multiput(941,229)(20.652,-2.065){0}{\usebox{\plotpoint}}
\put(957.95,227.23){\usebox{\plotpoint}}
\multiput(960,227)(20.756,0.000){0}{\usebox{\plotpoint}}
\put(978.70,227.00){\usebox{\plotpoint}}
\multiput(979,227)(20.652,-2.065){0}{\usebox{\plotpoint}}
\multiput(989,226)(20.756,0.000){0}{\usebox{\plotpoint}}
\put(999.40,225.86){\usebox{\plotpoint}}
\multiput(1008,225)(20.756,0.000){0}{\usebox{\plotpoint}}
\put(1020.09,224.66){\usebox{\plotpoint}}
\multiput(1026,224)(20.756,0.000){0}{\usebox{\plotpoint}}
\put(1040.81,224.00){\usebox{\plotpoint}}
\multiput(1045,224)(20.652,-2.065){0}{\usebox{\plotpoint}}
\put(1061.51,223.00){\usebox{\plotpoint}}
\multiput(1064,223)(20.756,0.000){0}{\usebox{\plotpoint}}
\put(1082.27,223.00){\usebox{\plotpoint}}
\multiput(1083,223)(20.652,-2.065){0}{\usebox{\plotpoint}}
\multiput(1093,222)(20.756,0.000){0}{\usebox{\plotpoint}}
\put(1102.97,222.00){\usebox{\plotpoint}}
\multiput(1112,222)(20.756,0.000){0}{\usebox{\plotpoint}}
\put(1123.71,221.70){\usebox{\plotpoint}}
\multiput(1130,221)(20.756,0.000){0}{\usebox{\plotpoint}}
\put(1144.43,221.00){\usebox{\plotpoint}}
\multiput(1149,221)(20.756,0.000){0}{\usebox{\plotpoint}}
\put(1165.19,221.00){\usebox{\plotpoint}}
\multiput(1168,221)(20.756,0.000){0}{\usebox{\plotpoint}}
\put(1185.89,220.12){\usebox{\plotpoint}}
\multiput(1187,220)(20.756,0.000){0}{\usebox{\plotpoint}}
\multiput(1197,220)(20.756,0.000){0}{\usebox{\plotpoint}}
\put(1206.64,220.00){\usebox{\plotpoint}}
\multiput(1215,220)(20.756,0.000){0}{\usebox{\plotpoint}}
\put(1227.40,220.00){\usebox{\plotpoint}}
\multiput(1234,220)(20.756,0.000){0}{\usebox{\plotpoint}}
\put(1248.15,220.00){\usebox{\plotpoint}}
\multiput(1253,220)(20.756,0.000){0}{\usebox{\plotpoint}}
\put(1268.87,219.35){\usebox{\plotpoint}}
\multiput(1272,219)(20.756,0.000){0}{\usebox{\plotpoint}}
\put(1289.61,219.00){\usebox{\plotpoint}}
\multiput(1291,219)(20.756,0.000){0}{\usebox{\plotpoint}}
\multiput(1301,219)(20.756,0.000){0}{\usebox{\plotpoint}}
\put(1310.36,219.00){\usebox{\plotpoint}}
\multiput(1319,219)(20.756,0.000){0}{\usebox{\plotpoint}}
\put(1331.12,219.00){\usebox{\plotpoint}}
\multiput(1338,219)(20.756,0.000){0}{\usebox{\plotpoint}}
\put(1351.87,219.00){\usebox{\plotpoint}}
\multiput(1357,219)(20.756,0.000){0}{\usebox{\plotpoint}}
\put(1372.63,219.00){\usebox{\plotpoint}}
\multiput(1376,219)(20.756,0.000){0}{\usebox{\plotpoint}}
\put(1393.39,219.00){\usebox{\plotpoint}}
\multiput(1395,219)(20.756,0.000){0}{\usebox{\plotpoint}}
\multiput(1405,219)(20.756,0.000){0}{\usebox{\plotpoint}}
\put(1414.14,219.00){\usebox{\plotpoint}}
\multiput(1423,219)(20.756,0.000){0}{\usebox{\plotpoint}}
\put(1433,219){\usebox{\plotpoint}}
\end{picture}

\vspace{5mm}
{\bf Fig. 2.}
\end{center}

\end{document}